\documentclass[a4paper,11pt]{article}
\usepackage{graphicx,amssymb,amstext,amsmath,mathabx,mathtools,esvect,xparse}
\usepackage{color,wrapfig}
\usepackage{floatflt}
\usepackage{lipsum,ragged2e}

\definecolor{cbl}{rgb}{0,0,1}               

\newcommand{\bc}{\begin{center}}
\newcommand{\ec}{\end{center}}
\def\ba#1{\begin{array}{#1}\displaystyle}
\newcommand{\ea}{\end{array}}

\newcommand{\beq}{\begin{equation}}
\newcommand{\eeq}{\end{equation}}
\newcommand{\beqa}{\begin{eqnarray}}
\newcommand{\eeqa}{\end{eqnarray}}

\newcommand{\n}{\nonumber\\}
\newcommand{\bi}{\begin{itemize}}
\newcommand{\ei}{\end{itemize}}

\newcommand{\bra}{\langle}
\newcommand{\ket}{\rangle}

\newcommand{\TT}{{\cal T}}
\newcommand{\tTT}{\tilde{\cal T}}

\usepackage{cite}

\definecolor{purple_nice}{rgb}{0.4,0.2,0.7}
\definecolor{fuel_blue}{RGB}{42,162,185}
\definecolor{YInMn_blue}{RGB}{46, 80, 144}
\definecolor{ultramarine}{RGB}{63, 0, 255}
\definecolor{KLEIN_blue}{rgb}{0, 0.18, 0.65}
\usepackage[linktocpage=true]{hyperref}
\hypersetup{
    colorlinks=true,
    linkcolor=YInMn_blue,
    citecolor=ultramarine,
    filecolor=fuel_blue,
    urlcolor=KLEIN_blue,
}

\makeatletter
\renewenvironment{abstract}{%
      \begin{center}%
        {\bfseries \normalsize\abstractname\vspace{\z@}}
      \end{center}%
      \quotation}
    {\endquotation}
\makeatother
\usepackage[normalem]{ulem}  

\begin{document}

\begin{titlepage}
\title{Symmetry Resolved Measures in Quantum Field Theory:\\ a Short Review}
\author{Olalla A. Castro-Alvaredo{\color{red} {$^\heartsuit$}} and Luc\'ia Santamar\'ia-Sanz{\color{blue} {$^\clubsuit$}}\\[0.3cm]}
\date{\small {\color{red} {$^\heartsuit$}}  Department of Mathematics, City, University of London, 10 Northampton Square EC1V 0HB, UK\\
\medskip
{\color{blue} {$^\clubsuit$}}  Department of Physics, University of Burgos, Plaza Misael Ba\~nuelos, Burgos, 09001, Spain \\
}
\maketitle
\begin{abstract}
In this short review we present the key definitions, ideas and techniques involved in the study of  symmetry resolved entanglement measures, with a focus on the symmetry resolved entanglement entropy. In order to be able to define such entanglement measures, it is essential that the theory under study possess an internal symmetry. Then, symmetry resolved entanglement measures quantify the contribution to a particular entanglement measure that can be associated to a chosen symmetry sector. Our review focuses on conformal (gapless/massless/critical) and integrable (gapped/massive) quantum field theories, where the leading computational technique employs symmetry fields known as (composite) branch point twist fields. 

\end{abstract}

\bigskip
\bigskip
\noindent {\bfseries Keywords:} Symmetry Resolved Entanglement Measures, Entanglement Entropy, Quantum Field Theory, Twist Fields, Integrability.

\vfill

\noindent 
{\Large {\color{red} {$^\heartsuit$}}}o.castro-alvaredo@city.ac.uk\\
{\Large {\color{blue} {$^\clubsuit$}}}lssanz@ubu.es\\

\hfill \today
\end{titlepage}
\section{Introduction}
The study of entanglement measures in the context of low-dimensional quantum field theory (QFT) has been a very active field of research within theoretical physics for the past 30 years. Among this class of models, 1+1D conformal field theories (CFTs), which capture the universal properties of quantum systems at criticality, have received the most attention. We find some of the most influential early works \cite{CallanW94,HolzheyLW94,Calabrese:2004eu} to be devoted precisely to the study of these models. The results of these early papers, in conjunction with numerical and analytical work in integrable spin chain models \cite{latorre1,Latorre2,Jin,latorre3}, that is, lattice versions of CFT and QFT in 1+1 dimensions, revealed how entanglement measures, such as the ubiquitous entanglement entropy \cite{bennett}, display universal scaling at conformal critical points. They also demonstrated numerically evaluating the entanglement entropy of an interval in a spin chain to be the most effective method for identifying and classifying critical points. 

In parallel with the developments above, which fall mainly within the areas of mathematical and theoretical physics, in the field of information theory the question has been asked of what constitutes {\it a good measure of entanglement}. A good review of possible answers to this question can be found in \cite{Plenio} and there are indeed many such answers: entanglement entropy, R\'enyi entropy, concurrence, fidelity, purity, negativity, to name just a few. It is natural to wonder why we may want to have so many distinct functions supposed to measure the same quantity. A partial answer is that different entanglement measures are suitable to different situations. In particular the nature of the quantum state (i.e. mixed or pure), the value of certain parameters (i.e. finite or zero temperature) and the nature of the partition (i.e. are we measuring the entanglement between complementary or non-complementary regions?), all play a role in the choice of entanglement measure. There are some very few properties that seem natural though: we would like good measures to increase in value when entanglement increases, to be invariant under certain transformations of the state and (in many cases) to be vanishing for unentangled regions or states and non-vanishing for entangled ones\footnote{This is not the case for instance for the negativity where the difference between so-called destillable and non-destillable entanglement plays a role and so there can be entangled states which have vanishing negativity \cite{destillable}.}. 

In recent years, particular attention has been paid to yet another feature of quantum systems, namely, whether or not an internal symmetry is present. Mathematically speaking, all measures of entanglement rely on the diagonalisation of a reduced density matrix. In the presence of a symmetry, the structure of this density matrix is altered, it becomes block-diagonal, with each block associated with a distinct symmetry sector. It is then natural to ask whether one could define measures of entanglement which capture not the full entanglement of the state but the contribution to it which can be associated to a chosen symmetry sector. This question has been recently answered in the affirmative and the aim of this review is to provide a summary of the main ideas, definitions, properties and techniques that play a role in the computation of {\it symmetry resolved entanglement measures} in the context of 1+1D QFT. 

 In the context of CFT, the definition of the symmetry resolved entanglement entropy (SREE) was put forward in \cite{GS}, where it was related to correlation functions of generalised (or composite) branch point twist fields. At around the same time, the role of symmetries and the contribution of symmetry sectors to the total entanglement was also studied in \cite{german3} for a quantum spin chain. In the context of entanglement measures, the connection with correlation functions of fields associated with conical singularities was introduced in \cite{Calabrese:2004eu}. A few years later, a different picture relating entanglement to correlators of symmetry fields associated with cyclic permutation symmetry in replica theories was introduced \cite{entropy,benext} which is the picture generalised in \cite{GS} to composite fields. The basic idea is that in theories that possess an underlying symmetry (common symmetries in QFT are $U(1)$ and $\mathbb{Z}_k$ for some integer $k$) entanglement can be expressed as a sum over contributions from different symmetry sectors. Remarkably such contributions are experimentally measurable \cite{Islam,expSRE1,expSRE2,expSRE3,expSRE4,Lukin_2019}, which provides further motivation to study this quantity.  

\medskip
This review article is organised as follows: in Section \ref{definitions} we present the definitions of symmetry resolved entanglement entropies and partition functions, as well as introducing the charged moments. In Section \ref{BPTFsec} we review the properties of (composite) twist fields and their applications in the context of entanglement. We present two examples: the sine-Gordon model and the Ising field theory. Section \ref{review} contains a literature review. In Section \ref{scaling} we derive the main properties of the symmetry resolved entanglement entropy, including equipartition, and show that they follow from simple CFT and QFT arguments. In Section \ref{FFP} we introduce the form factor program for composite twist fields.  In Section \ref{corrections} we show how finite region size corrections can be derived using form factors. In Section \ref{excitedstates} we discuss the universal properties of the symmetry resolved entanglement entropy of a certain class of excited states. We conclude in Section \ref{conclu}.
 
 \section{Symmetry Resolved Entanglement Entropy: Definitions}
 \label{definitions}
 Let us now introduce some basic definitions and notation.  This being a short review, we will consider the simplest setup. This means that we will discuss only the case of spacial bipartitions of pure states and study a single symmetry resolved measure, namely the entropy. Let $|\Psi\ket$ be a pure state of a 1+1D QFT and let us define a bipartition of space into two complementary regions $A$ and $\bar{A}$ so that the Hilbert space of the theory $\mathcal{H}$ also decomposes into a direct product $\mathcal{H}_A \otimes \mathcal{H}_{\bar{A}}$. Then the reduced density matrix associated to subsystem $A$ is obtained by tracing out the degrees of freedom of subsystem $\bar{A}$ as
\beq
\rho_A=\mathrm{Tr}_{\bar{A}}(|\Psi\ket \bra \Psi|)\,,
\eeq
and the von Neumann (or entanglement) and $n$th R\'enyi entropy of a subsystem $A$ are defined as
\beq
S=-\mathrm{Tr}_A(\rho_A \log \rho_A)\quad \mathrm{and} \quad S_n=\frac{\log(\mathrm{Tr}_A \rho_A^n)}{1-n}\,,
\label{SS}
\eeq
where $\mathrm{Tr}_A \rho_A^n:={\mathcal{Z}}_n/{\mathcal{Z}}_1^n$ can be interpreted as the normalised partition function of a theory constructed from $n$ non-interacting copies or replicas of the original model. We will call this the ``replica theory"\footnote{For readers more familiar with discrete systems, such as quantum spin chains, it would be more natural to think of the state $|\Psi\ket$ as a state of the full chain and the region $A$ as a subset of the spins in the chain, with $\bar{A}$ its complement. The QFT description is an appropriate continuous version of the spin chain, where the number of spins becomes very large and their mutual spacing very small, in a controlled way so that the fundamental degrees of freedom become local quantum fields rather than local spins.}. It is easy to show that $S=\lim_{n\rightarrow 1} S_n$.
\medskip

 In the presence of an internal symmetry, we can also define a symmetry operator $Q$ and its projection onto subsystem $A$, $Q_A$. In the original work \cite{GS}, $Q_A$ was specifically related to the number of charged particles in one region when inserting a space-time Aharonov-Bohm flux into the $n$-sheeted space, which couples to the particles' charge. They assume that the pure wavefunction $|\Psi\ket$ of the total system is an eigenfunction of the total conserved quantity, which is a sum of contributions of the two subsystems. This hypothesis can be justified because the total particle number in a system is usually fixed, so the conserved quantity associated to the particle number and the Hamiltonian commute, and share a basis of common eigenstates. This property extends to other types of symmetry too, ensuring that $[Q_A, \rho_A]=0$. If $q$ is the eigenvalue of operator $Q_A$ in a particular symmetry sector, then ${\mathcal{Z}}_n(q)=\mathrm{Tr}_A(\rho_A^n \mathbb{P}(q))$ with $\mathbb{P}(q)$ the projector onto the symmetry sector of charge $q$, can be identified as the {\it symmetry resolved partition function}. In terms of this object, the symmetry resolved entanglement entropies (SREEs) can be written as
\beq
S_n(q)=\frac{1}{1-n} \log\frac{{\mathcal{Z}}_n(q)}{{\mathcal{Z}}_1^n(q)}\quad \mathrm{and} \quad S(q)=\lim_{n\rightarrow 1} S_n(q)\,.
\label{sym}
\eeq
An interesting feature of these formulae is that if we compute the total entropy $S$ in terms of the symmetry resolved contributions $S(q)$ we find the following structure
\beq 
S=\sum_q \left[p(q) S(q)- p(q) \log p(q)\right]\,,
\label{number}
\eeq 
where $p(q):=\mathcal{Z}_1(q)$ represents the probability that a measurement of the symmetry charge delivers the value $q$. In the context of symmetry resolved measures, the first contribution is known as {\it configuration entropy} which, as we can see, is the sum of the SREEs weighted by the probability of the corresponding charge sector. The second contribution is the {\it number entropy} or {\it fluctuation entropy}, which is associated to fluctuations in the value of the charge of subsystem $A$. If particle number is the conserved quantity in the system, then the number entropy relates the number of the particles in one subsystem to that of the other, for a given total particle number, that is, it characterises particle-number fluctuations between subsystems and is experimentally measurable \cite{Lukin_2019, Wiseman}. Thus, the total entropy is not just the sum over symmetry sectors of the symmetry resolved entropy. Instead we have a contribution of this type, but with each SREE contribution $S(q)$ weighted by $p(q)$, plus the number entropy\footnote{The names ``number entropy" and ``configuration entropy" may seem counterintutitive since in the context of statistical mechanics, the quantity $-k_B \sum_n p_n \log p_n$, where $k_B$ is the Boltzmann constant and $p_n$ is the probability of state $n$, is called the configuration or configurational entropy.}.

As discussed in \cite{GS} the symmetry resolved partition function can be obtained from its Fourier modes, the so-called {\it charged moments} $Z_n(\alpha)=\mathrm{Tr}_A(\rho_A^n e^{2\pi i\alpha Q_A})$. The expression for these depends on the type of symmetry under consideration, in particular whether it is continuous or discrete. Two common examples are 
\beq
\mathcal{Z}_n(q)  = \begin{cases}\int\limits^{\frac{1}{2}}_{-\frac{1}{2}} \frac{d\alpha}{2\pi}Z_n(\alpha) e^{-2\pi i\alpha q},  \qquad \qquad\text{for a} \,\, U(1) \,\, \text{continuous \,symmetry},\\
\frac{1}{N}\sum\limits^{N-1}_{k=0} Z_n\left( \frac{2\pi k}{N}\right) e^{-i \frac{2\pi k q}{N}},  \,\, \qquad \text{for a} \,\, \mathbb{Z}_N \, \,\text{discrete\,symmetry}\,.\end{cases}
\label{partition}
\eeq
The key contribution of \cite{GS} was the realisation that these charged moments can be expressed in terms of correlation functions of symmetry fields, and therefore are more easily accessible analytically than the partition functions themselves. These correlation functions are directly proportional to the partition function of the $n$-sheeted Riemann surface with a generalised Aharonov-Bohm flux. We now proceed to introduce branch point twist fields and their composite versions.

\section{Branch Point Twist Fields in Quantum Field Theory}
\label{BPTFsec}
Consider again the formulae (\ref{SS}). It is clear that if we know how to compute ${\rm Tr_A \rho_A^n}$ we will then have access to all entropies. Indeed, research into entanglement measures in QFT often centers around the numerical and/or analytical evaluation of this quantity.  
Many investigations of entanglement in QFT are based on numerical simulations of quantum spin chains in the scaling limit. This includes some pioneering works such as \cite{latorre1,Latorre2,Jin,latorre3}.  In this setting, the reduced density matrix $\rho_A$ is large but finite so that a suitable diagonalisation method can be found. Once all eigenvalues are know, the evaluation of the trace is straightforward. There are relatively few models for which analytic computations of such a complex quantity are possible. 
\begin{wrapfigure}{l}{0.45\textwidth}
  \begin{center}
    \includegraphics[width=0.45\textwidth]{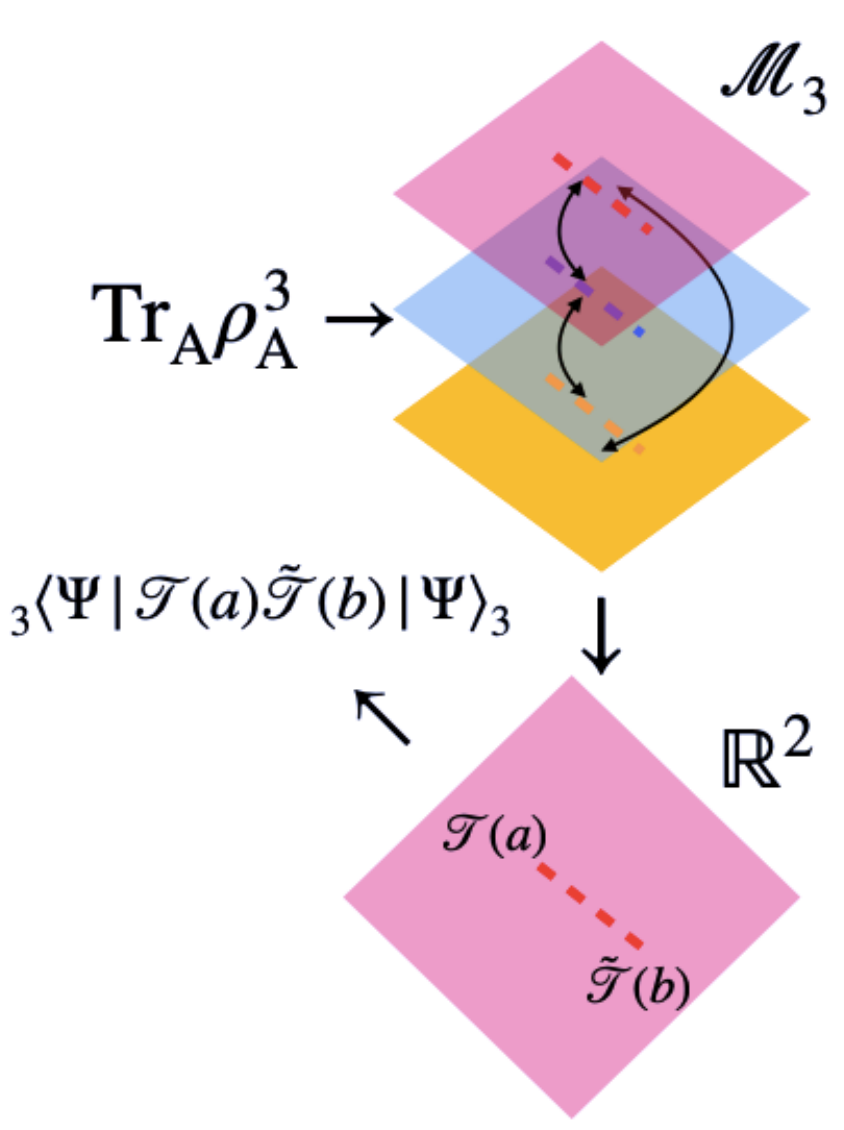}
  \end{center}
  \caption{Relationship between the replica partition function and the correlation function of branch point twist fields for $n=3$ and a compact region of length $|a-b|$. $|\Psi\ket_n$ is a pure state in the replica theory.}
  \label{sheets}
\end{wrapfigure}
Exceptions to this are free models such as the XY chain where the reduced density matrix can be explicitly constructed and diagonalised both at and away from equilibrium \cite{Jin,FC}, and CFTs, where many leading features of entanglement are well-understood also both at equilibrium \cite{HolzheyLW94,Calabrese:2004eu,Calabrese:2005in} and away \cite{quench, quench2,quench3, quenchesCC, AC, AC2}. An scarcity of analytical results  is typical when going beyond criticality. However, over the past few years, analytical results for interacting QFTs, especially integrable ones have become accessible through the use of branch point twist fields (BPTFs). It can be shown that the trace ${\rm Tr_A \rho_A^n}$ is proportional to a correlation function of BPTFs and a suitable generalisation of this map has also been found for symmetry resolved measures. 

The trace $\rm{Tr}_A\rho_A^n$ associated with a connected subsystem $A$ of length $|a-b|$ is identified with a (normalised) partition function $\mathcal{Z}_n/\mathcal{Z}_1^n$ on an $n$-sheeted Riemann manifold $\mathcal{M}_n$. This manifold is exactly the Riemann surface of the complex function $\sqrt[n]{\frac{z-a}{z-b}}$ with $a, b$ the branch points. This connectivity of the manifold originates from the structure of the $n$th power (leading to $n$ connected sheets) and the trace (leading to cyclically connected sheets). Then, the problem of computing this partition function can be mapped to the problem of computing a correlation function in the complex plane with two BPTF insertions at the branch points, see Fig.\ref{sheets}. 

The geometric complexity of the Riemann surface is then encoded into the exchange relations of the BPTF with other local fields of the theory. These branch point twist fields are twist fields in the standard sense in QFT, that is, they are symmetry fields associated to an internal symmetry of the theory, much like the order and disorder fields in the Ising field theory \cite{BaBe,BeLe,BeLe0,Yurov} which are associated to $\mathbb{Z}_2$ symmetry. The fields $\TT, \tilde{\TT}$ are associated also to a discrete symmetry, namely the two opposite cyclic permutation symmetries $\sigma: i\mapsto i+1$ and $\sigma^{-1}: i+1\mapsto i$  ($i=1,\ldots,n,\;n+i\equiv i$) present in the replica theory
\beqa &&
    \TT=\TT_\sigma~,\quad\quad\, \sigma\;:\; i\mapsto i+1 \ {\rm mod} \,n\,, \n &&
    \tilde{\TT}=\TT_{\sigma^{-1}}~,\quad \sigma^{-1}\;:\; i+1\mapsto i \ {\rm mod} \,n\,.
\eeqa
Thus, given a QFT containing a local field $\varphi(\bf{x})$ with coordinates $\bf{x}:=(x^0,x^1)$, its replica version contains $n$ fields $\varphi_i(\bf{x})$ and their equal-time exchange relations with the BPTF and its hermitian conjugate are as follows:
\begin{eqnarray}
\begin{array}{ll}
    \varphi_{i}(\bf{x})\mathcal{T}(\bf{y}) = \mathcal{T}(\bf{y}) \varphi_{i+1} (\bf{x}),  &  \qquad y^{1}> x^{1},\\
      \varphi_{i}(\bf{x})\mathcal{T}(\bf{y}) = \mathcal{T}(\bf{y}) \varphi_{i}(\bf{x}), &  \qquad x^{1}>y^{1},\\
      \varphi_{i}(\bf{x})\tTT(\bf{y}) = \tTT(\bf{y}) \varphi_{i-1}(\bf{x}), & \qquad y^{1}> x^{1},\\
      \varphi_{i}(\bf{x})\tTT(\bf{y}) = \tTT(\bf{y}) \varphi_{i}(\bf{x}), & \qquad x^{1}> y^{1}. 
\end{array}
    \label{cr}
\end{eqnarray}
At this stage, it is worth presenting a brief history of the fields $\TT, \tTT$. In the context of entanglement, their description as symmetry fields in replica theories, appeared first in \cite{entropy}. A connection between entanglement measures and correlation functions of conical fields in CFT was first proposed in \cite{Calabrese:2004eu}. However, BPTFs were described much earlier in a different context. They emerged in the study of orbifold CFT, and their conformal dimensions were first obtained in \cite{Kniz,orbifold}. They are given by
\beq
\Delta_n=\frac{c}{24}\left(n-\frac{1}{n}\right)\,,
\eeq
and are functions of the central charge $c$ and the replica number $n$.
\subsection{Composite Fields}
Returning now to the SREE, we recall that the basic building blocks are the moments $Z_n(\alpha)$ (let us assume $U(1)$ symmetry for now). In \cite{GS} it was shown that these moments can also be expressed in terms of correlators of symmetry fields. These symmetry fields should implement cyclic permutation symmetry and the internal symmetry of the theory {\it simultaneously} so they are still BPTFs but not the fields $\TT$, $\tTT$ described in the previous subsection. 

The problem of how such fields may be defined is easy to solve in CFT. They must be fields that are formed by {\it composition} of two twist fields, one associated to cyclic permutation symmetry, that is $\TT$, and one associated to the internal symmetry of the theory, which would vary from theory to theory. Note that composition with local (non twist) fields can also be considered, as done in \cite{CDL,Levi} and this is of interest in the context of the entanglement entropy of non-unitary QFTs \cite{BCDLR,bcd15}. We will call these fields, composite twist fields (CTFs).

It is instructive to present the following definition of a CTF  $:\!\mathcal{T}\,\phi\!:$, following \cite{CDL}.  Let $\phi$ 
be a local field in a CFT, then the CTF in  the replica theory can be defined as
\begin{equation}
:\!\mathcal{T}\,\phi\!:\!(y)\,:= n^{2\Delta -1}\lim_{x\rightarrow y} |x-y|^{2\Delta \left(1-\frac{1}{n}\right)} \sum_{j=1}^n \mathcal{T}(y) \phi_j(x)\,,
\label{ccft}
\end{equation}
where $\phi_j(x)$ is the copy of field $\phi(x)$ living in replica $j$, $:\quad:$ represents normal ordering and the power law, involving the conformal dimension of the field $\phi$, denoted by $\Delta$, is obtained using conformal symmetry. The CTF is then the leading field in the operator product expansion of $\TT(y)$ with $\sum_j \phi_j(x)$. The prefactor $ n^{2\Delta -1}$ ensures conformal normalisation of the two-point function of CTFs.  This definition is then extended to more general QFTs (i.e. non conformal) in the usual way, by seeing the CTF as the off-critical versions of its conformal counterpart (\ref{ccft}). Employing once more conformal arguments, it is possible to show that the conformal dimension of the CTF is given by \cite{CDL,GS}
\beq 
\Delta_n^\phi=\Delta_n+\frac{\Delta}{n}\,.
\label{dim}
\eeq 
In the context of symmetry resolution a special choice of the field $\phi$ is made. Namely, we now need this field to also be a symmetry field. The nature of the field will depend on the theory. Here we provide two examples: the case of a continuous $U(1)$ symmetry, as found in the sine-Gordon model \cite{Zamolodchikov:1977yy}, and the case of a discrete $\mathbb{Z}_2$ symmetry, as found in the Ising field theory \cite{BaBe}. The first is an interacting integrable massive QFT while the second is a free massive QFT. Here `massive' is understood as non-critical, that is, in both cases there is a mass gap. The use of CTFs ties up with the Riemann surface picture of Fig.~1 through an argument that was first presented in \cite{GS}. The insertion of the additional twist field $\phi$ can be seen as introducing an Aharonov-Bohm flux on one of the Riemann sheets \footnote{From the viewpoint of computing entanglement measures employing form factors of CTFs this flux can be spread over all the the Riemann sheets, as long as it combines to its total value. This idea was employed in \cite{ourPartIII} and is represented by Fig.~1 in \cite{BCD}.}. In terms of these fields  we have that the charged moments $Z_n(\alpha)$ for a connected region of length $\ell$ can be written as two-point functions of the CTF and its conjugate. We will see some examples later.

\subsubsection{The sine-Gordon Model}
\label{SG}
The sine-Gordon action is given in terms of a fundamental bosonic field $\varphi$ and coupling constants $\gamma, g$ as:
\beq 
\mathcal{A}=\int dx dt \left[\frac{1}{16\pi} [(\partial_0 \varphi)^2- (\partial_1 \varphi)^2] -2\gamma \cos(g \varphi)\right]\,,
\label{sgL}
\eeq 
where $\gamma$ is related to the total kink energy, sometimes called the classical kink mass. The model has $U(1)$ symmetry, seen by the invariance of $\mathcal{A}$ under the shift $\varphi \mapsto \varphi+ (2\pi/g)$. Famously, the model has topological sectors associated with different $U(1)$ charges. Field configurations in different topological sectors cannot evolve into each other without violating the finite energy condition. Hence, the conserved topological indices in the theory come from the finite energy condition and not from a continuous symmetry. At classical level, the solutions to the field equation interpolate between different vacua associated with those sectors, thus (in some cases) carrying a topological charge. There are three types of fundamental solutions, known as soliton, antisoliton and breathers. The latter can be seen as bound states of the former which ``breath" in the sense that they are time-dependent solutions. As we see in Section \ref{FFP} these classical solutions are promoted to stable quantum excitations in the quantum model \cite{solitonbook,Zamolodchikov:1977yy}. 

The twist field  associated with $U(1)$ symmetry is the simple vertex operator
\begin{equation}
\mathcal{V}_{\alpha}=\exp\left({\frac{i\alpha g \varphi}{2\pi}}\right)\,,\label{SGU(1)Field}
\end{equation}
with conformal dimension $\Delta=(g^2\alpha^2)/(32\pi^3)$. The short-distance limit of the model is a massless free boson with central charge $c=1$. 
The exchange relations of this $U(1)$ twist field with other local fields in the theory are characterised by 
the semi-locality (or mutual locality) 
index $e^{i \kappa \alpha}$  \cite{BaBe,Yurov} via the equal-time
exchange relations
\beq
\label{eq:LocalityDefBIS}
\mathcal{V}_{\alpha}({\bf x})\phi_{\kappa}({\bf y})=\left\{ \begin{array}{cc} e^{i\kappa\alpha}\phi_{\kappa}({\bf y})\mathcal{V}_{\alpha}({\bf x}), & \quad y^{1}>x^{1},\\
 \phi_{\kappa}({\bf y})\mathcal{V}_{\alpha}({\bf x}), & \quad x^{1}>y^{1},\end{array}\right. 
\eeq
with $\kappa=\pm,0$. The fields $\phi_{\pm,0}$ are
associated with the creation of a soliton ($+$), an antisoliton
$(-)$ or a neutral particle $(0)$ (usually called a breather). 
The mutual locality factor $e^{i \kappa \alpha}$ and its physical
meaning are in agreement with the intuitive picture presented in \cite{GS}, where it is associated with the
insertion of the Aharonov-Bohm flux on one of the Riemann sheets.
Consequently, the $U(1)$ CTF denoted as $\mathcal{T}^{\alpha}_n({\bf x})$
can be understood formally as $:\mathcal{T}\mathcal{V}_{\alpha}:({\bf x})$ in the sense of (\ref{ccft}), and in a replica theory, is characterised by equal-time exchange relations 
\beq
\mathcal{T}_n^{\alpha}({\bf x})\mathcal{O}_{p,i}({\bf y}) =\left\{ \begin{array}{cc}
e^{\frac{ip\alpha}{n}}\mathcal{O}_{p,i+1}({\bf y})\mathcal{T}_n^{\alpha}({\bf x}),& \quad y^{1}>x^{1},\\
 \mathcal{O}_{p,i}({\bf y})\mathcal{T}_n^{\alpha}({\bf x}),& \quad x^{1}>y^{1},
 \end{array}\right. 
\label{U(1)CBPTFSpatialExchange}
\eeq
with respect to quantum fields $\mathcal{O}_{p,i}$ living on the
$i$th replica and possessing $U(1)$ charge $p\in\mathbb{Z}$.
Similarly, 
\beq
\tilde{\mathcal{T}}_n^{\alpha}({\bf x})\mathcal{O}_{p,i}({\bf y}) = \left\{ \begin{array}{cc} 
e^{-\frac{ip\alpha}{n}}\mathcal{O}_{p,i-1}({\bf y})\tilde{\mathcal{T}}_n^{\alpha}({\bf x}), & \quad  y^{1}>x^{1},\\
 \mathcal{O}_{p,i}({\bf y})\tilde{\mathcal{T}}_n^{\alpha}({\bf x}),&  \quad x^{1}>y^{1}.
 \end{array}
 \right.
 \label{U(1)CBPTFSpatialExchange(AntiTwistField)}
\eeq
The choice of the phases $\pm \alpha/n$ is motivated by
requiring that the total phase picked up by a charged particle
(associated with a unity of charge) is $e^{\pm i\alpha}$ when
turning around each of the branch points.
\subsubsection{The Ising Field Theory}
\label{Ising}
The Ising field theory describes a free Majorana fermion $\psi(x)$ of mass $m$ with action
\beq 
\mathcal{A}=\frac{1}{2\pi} \int dx dt \left[\psi \bar\partial \psi + \bar\psi\partial \bar\psi + i m \bar\psi\psi \right]\,,
\label{AIsing}
\eeq 
It is well-known that the Ising field theory has an internal $\mathbb{Z}_2$ symmetry. This can be seen from the invariance of the action under $\psi \mapsto -\psi$. Associated to this symmetry there are two twist fields: $\sigma$, the spin field (order operator), and $\mu$ (disorder operator)\footnote{In this paper we use the conventions of \cite{Yurov}, which corresponds to choosing the disordered phase of the model, where the fields $\sigma (\mu)$ are odd (even) with respect to the Majorana fermion $\psi$.}. The theory also contains a free Majorana fermion field $\psi$ so that, in the disordered phase of the theory, the three fields can be characterised by their mutual equal-time exchange relations \cite{BaBe,BeLe,BeLe0,Yurov}:
\beq
\psi ({\bf x})\sigma({\bf y})=\left\{ \begin{array}{cc}
\sigma({\bf y}) \psi({\bf x}), & \quad y^1>x^1,\\
\sigma ({\bf y}) \psi({\bf x}), & \quad x^1>y^1,
\end{array}\right. \quad \mathrm{and}\quad \psi ({\bf x})\mu({\bf y})=\left\{ \begin{array}{cc}
-\mu({\bf y}) \psi({\bf x}),& \quad y^1>x^1,\\
\mu ({\bf y}) \psi({\bf x}), & \quad x^1>y^1.
\end{array}\right.
\label{simu}
\eeq 
In the conformal limit, the fields $\mu, \sigma$ both have dimension $\Delta=1/16$ and the theory can be seen as the massive perturbation of a massless free fermion with central charge $c=1/2$.
In a replica theory, the Majorana fermion is labelled by a copy index $\psi_j({\bf x)}$ and satisfies the usual exchange relations (\ref{cr}) with the BPTF.  The leading fields in the operator product expansions of $\TT$ with $\sum_j \sigma_j$ and $\sum_j \mu_j$, here denoted by $\TT_n^\pm$ respectively, are CTFs satisfying the exchange relations
\beq
\psi_j ({\bf x})\TT_n^\pm({\bf y})=\left\{ \begin{array}{cc}
\pm \TT_n^\pm({\bf y}) \psi_{j+1}({\bf x}),& \quad y^1>x^1,\\
\TT_n^\pm ({\bf y}) \psi_j({\bf x}), & \quad y^1<x^1,
\end{array}\right.
\eeq
and 
\beq 
 \psi_j ({\bf x})\tTT_n^\pm({\bf y})=\left\{ \begin{array}{cc}
\pm \tTT_n^\pm({\bf y}) \psi_{j-1}({\bf x}),& \quad y^1>x^1,\\
\tTT_n^\pm ({\bf y}) \psi_j({\bf x}), & \quad y^1<x^1.
\end{array}\right.
\label{Tmu}
\eeq 
\section{A Brief Literature Review}
\label{review}
Starting from these basic ideas, symmetry resolved measures have been investigated for many models. In fact, the idea that we might ``resolve" entanglement measures according to some property already appeared in early works such as \cite{Laflorencie_2014}, where spin resolution is considered. Many studies of symmetry resolved entanglement measures, starting with the original works \cite{GS,german3}, have considered either 1+1D CFTs or critical spin chains because of the specific analytic and numerical methods that are available in those cases. While \cite{Bonsignori_2019,Capizzi_2020} studied the SREE of massless free fermions and of excited states of CFT, respectively,  \cite{Bonsignori:2020laa} considered the effect of boundaries,  and \cite{Estienne:2020txv,di2023boundary} studied finite size corrections, also in CFT. Specific CFTs characterised by intricate mathematical structures and physical applications, such as the Wess-Zumino-Novikov-Witten model \cite{Calabrese:2021wvi}, and theories possessing  categorical non-invertible symmetries \cite{Sierra_2024} continue to be intensively studied at present.

The SREE is just one among a large family of entanglement measures, all of which can be subjected to symmetry resolution. Studies of other measures also abound. For example, \cite{Cornfeld:2018wbg,Murciano:2021djk} studied the symmetry resolved negativity of massless free fermions, while in \cite{Chen:2021pls,Capizzi:2021zga} the relative entropies of CFT were given the symmetry resolved treatment. Likewise for the symmetry resolved fidelities of gaussian states in \cite{Parez:2022sgc}, the Page curve \cite{Murciano:2022lsw} and the CCNR negativity \cite{Yin:2022toc,Berthiere:2023gkx,Bruno:2023tez}\footnote{CCNR stands for computable cross-norm or realignment.}. In particular, in cases when the entangling region consists of disconnected parts, it is useful to compute the so-called multi-charged moments, with potentially distinct charges for each interval. These have been studied for instance in \cite{Parez:2021pgq,Ares:2022gjb,Gaur:2023yru} for the massless Dirac field theory and the compactified free boson, respectively.

Free theories, whether critical or not, provide fertile ground for the study of entanglement measures. This is because there is a wider range of analytical methods at our disposal, compared to interacting models.  For example, for free fermionic spin chains, one may compute entanglement measures, including symmetry resolved ones, by employing the properties of Toeplitz determinants, as exploited in \cite{Bonsignori:2020laa,Ares:2022gjb,Ares:2022hdh}. Toeplitz determinants emerge naturally when computing the reduced density matrix of free fermionic systems\footnote{Essentially, the non-vanishing entries of the reduced density matrix are two-point correlators which can be expressed in terms of Toeplitz determinants \cite{Jin}.}. The Dirac free fermion, which is a free massive QFT with $U(1)$ symmetry, has also been studied in \cite{Horvath:2020vzs,Horvath:2021fks,CTFMichele} employing the form factor program (see Section \ref{FFP}).

The form factor program is the most successful analytic method for dealing with integrable 1+1D QFT, and can be employed both for interacting and free theories. It is a systematic approach to computing matrix elements of local and twist fields which can then be used as building blocks for correlation functions. Given the relationship between correlation functions of BPTFs and CTFs and entanglement measures it is no surprise that the form factor program has been employed to study entanglement measures in QFT. In the context of symmetry resolution, one needs to first extend the program to deal with CTFs, a task that was completed in \cite{Horvath:2020vzs}. The form factor equations proposed in that work have subsequently been employed to study the SREE in the ground state for the sine-Gordon \cite{Horvath:2021rjd} and Potts models \cite{Potts}, as well as for excited states in free massive fermions and bosons \cite{ourPartI,ourPartII}, with a subsequent extension to studying the negativity in free fermion theories \cite{ourPartIII}. An interesting recent work has used the form factor technique to investigate massless flows between two CFTs in the context of symmetry resolution \cite{Rottoli:2023jvw}. Form factors are however not the only way to tackle interacting theories. There are of course numerical methods, which are also employed in many of the papers cited above. For integrable lattice systems we may also employ the corner transfer matrix approach, which is particularly well suited to evaluating entanglement measures in the half-infinite line. This approach has also been employed for symmetry resolved measures in \cite{Murciano:2019wdl}.

A further viewpoint on entanglement in QFT is provided by the holographic picture stemming from string theory. It has been known since the pioneering work \cite{Ryu:2006bv,Ryu:2006ef} that there is a quantitative relationship between the entanglement entropy of a CFT and the geometry of an associated AdS spacetime. More precisely the entanglement entropy of a region $A$ where a CFT is defined, is proportional to the area of a ``minimal surface" whose boundary is $A$. Symmetry resolved measures have also been studied within this holographic setting  \cite{Belin:2013uta, Caputa:2015qbk,Zhao:2020qmn,Weisenberger:2021eby,Baiguera:2022sao,Zhao:2022wnp}.

As mentioned at the beginning of this section, a lot of results for entanglement measures in general, and symmetry resolved ones in particular, consider lattice models. We have already mentioned some of these works \cite{Laflorencie_2014,german3,FG,Bonsignori:2020laa,Bonsignori_2019,Murciano:2019wdl} when discussing free theories earlier. Additional examples are found in 
\cite{Fraenkel:2019ykl,Murciano:2020lqq}, and in \cite{barghathi2018renyi,Barghathi:2019oxr,Tan_2020}. In the latter, there is also an interesting focus on the role played by different contributions to the total entanglement as seen by equation (\ref{number}). It is found that, depending on the symmetry of the system or selection rules (e.g. fixing the particle number), only a certain contribution to the entanglement entropy will be ``operationally accessible", that is the number entropy introduced after equation (\ref{number}). 

Having considered different partitions, theories, states and measures, it is also natural to venture beyond equilibrium states and on to investigating out-of-equilibrium protocols \cite{quench,quench2,quench3,quenchesCC}. These have been extensively studied for the past 10 years, and there is a vast amount of literature on the subject (see e.g.~the reviews \cite{ESSLER_ghd_review,doyon_GHD_review,Bernard:2016nci} and the book \cite{zwanzig2001nonequilibrium}). Suffice to say that any ideas and techniques employed for standard entanglement measures, say, the quasi-particle picture \cite{AC,AC2}, can be applied to studying symmetry resolved measures too. Some examples are found in \cite{FG, Caputa:2013eka, Tan_2020,Parez:2020vsp,expSRE3,Fraenkel:2021ijv,Bonsignori:2019naz,Parez:2021pgq,Parez:2022sgc,chen2023energy,Parez:2022xur,Scopa_2022}.

Finally, let us mention studies concerned with other types of theories, that is models that are neither critical in the standard sense nor integrable.  For example, the symmetry resolved entanglement of quantum spin chains with random couplings has distinct properties, which are different from but resemble those of CFT, and was studied in \cite{Turkeshi:2020yxd}. The symmetry resolved entanglement in many-body localised systems has been studied in \cite{Kiefer_Emmanouilidis_2020,Kiefer_Emmanouilidis_2021}. It is also possible to examine topological phases \cite{Monkman:2020ycn,Azses:2020wfx,Monkman:2023hup,Monk}, topological defects \cite{Horvath:2023xoh} and more general statistics, like that of anyons \cite{Cornfeld:2018wtp} through the lens of symmetry resolution. Recently, the SREE of non-local QFTs has been studied in \cite{nonlocalqft}.

\section{Scaling and Equipartition}
\label{scaling}
What are the most salient properties of the SREE?
The answer to this question appeared already in the early works \cite{GS,german3} and has been further investigated in many of the publications listed above. Three main properties were found:
\begin{enumerate}
\item In a critical system, the SREE scales logarithmically with the size of the subsystem, just as the total entropy does.
\item In a critical system, the SREE exhibits the property of {\it equipartition}, namely in some limit at least, its value is independent of the chosen symmetry sector.
\item In a critical system, the SREE exhibits a double-logarithmic correction which contains information about the scaling dimension of the symmetry field.
\end{enumerate}
All of these properties can be easily shown using standard CFT arguments. In particular, we may employ the CTFs introduced earlier, together with the information about their scaling dimension, and the fact that they are primary fields whose correlators are normalised in the usual way. In order to perform an explicitly computation, we will consider the case of the $U(1)$ symmetry in sine-Gordon with the definitions given in subsection \ref{SG}. We have said earlier that the charged moments are defined in terms of correlators of these fields. If we are computing the symmetry resolved entropies for a continuous interval of length $\ell$ in the ground state, we can write that
\beq 
Z_n(\alpha)=\varepsilon^{4\Delta_n^\alpha} {}_n\bra 0|\TT_n^\alpha(0)\tTT_n^\alpha(\ell)|0\ket_n \,,
\label{charge}
\eeq 
where $\Delta_n^\alpha$ is the dimension of the CTF and $\varepsilon$ is a non-universal short-distance cut-off\footnote{In spin chains $\varepsilon$ is proportional to the lattice spacing. The introduction of the cut-off $\varepsilon$ allows for the comparison between QFT and lattice results. See \cite{entropy} for an explicit computation in the Ising model where the exact relationship between the cut-off $\varepsilon$ and the lattice spacing was found.}, in this case
\beq 
\Delta_n^\alpha=\frac{1}{24}\left(n-\frac{1}{n}\right)+ \frac{g^2\alpha^2}{4(2\pi)^3 n}.
\eeq 
Note that, with this notation, $\Delta_n^0=\Delta_n$ is the dimension of the BPTF and $\Delta_1^\alpha$ is the dimension of the $U(1)$ field (which we called $\Delta$ in the general formula (\ref{dim})). 
In a CFT, with the normalisation (\ref{ccft}), we have  that the two-point function scales with the distance between fields in the usual way
\beq 
Z_n(\alpha)=\left(\frac{\varepsilon}{\ell}\right)^{4\Delta_n^\alpha}\,,
\label{MCFT}
\eeq 
thus, all the $\alpha$-dependence comes from the scaling dimension of the composite twist field. In order to obtain the symmetry resolved partition function, we need to Fourier-transform this expression with respect to the variable $\alpha$ using the formula (\ref{partition}):
\beq
\mathcal{Z}_n(q)=\int_{-\frac{1}{2}}^{\frac{1}{2}} \frac{d\alpha}{2\pi} \left(\frac{\varepsilon}{\ell}\right)^{4\Delta_n^\alpha} e^{-2\pi i \alpha q}=\left(\frac{\varepsilon}{\ell}\right)^{4\Delta_n^0}\int_{-\frac{1}{2}}^{\frac{1}{2}} \frac{d\alpha}{2\pi} \left(\frac{\varepsilon}{\ell}\right)^{\frac{g^2 \alpha^2}{(2\pi)^3 n}} e^{-2\pi i \alpha q}\,.
\eeq 
The integral can be computed exactly. We note as well that $\left(\varepsilon/\ell\right)^{4\Delta_n^0}$ is nothing but the partition function $\mathcal{Z}_n$ (non-symmetry resolved) so that, from this term, we will subsequently recover a leading contribution to the SREE that is identical to the full entropy.  For $\varepsilon/\ell<1$ we have that $\log (\ell/\varepsilon)>0$. In that case we can rewrite the integral and compute it exactly to obtain
\begin{eqnarray}
\mathcal{Z}_n(q) &=& \left(\frac{\varepsilon}{\ell}\right)^{4\Delta_n^0}\int_{-\frac{1}{2}}^{\frac{1}{2}} \frac{d\alpha}{2\pi} e^{- \alpha^2 \frac{\tilde{\Delta}}{n} \log \frac{\ell}{\varepsilon}} e^{-2\pi i \alpha q}
\nonumber\\
& = & \frac{1}{2\pi} \left(\frac{\varepsilon}{\ell}\right)^{4\Delta_n^0} 
\frac{\sqrt{\pi} e^{-\frac{\pi^2 q^2}{\frac{\tilde{\Delta}}{n}  \log\frac{\ell}{\varepsilon}}}}{2\sqrt{\frac{\tilde{\Delta}}{n}  \log\frac{\ell}{\varepsilon}}}\left({\rm{Erf}}\left(\frac{\frac{\tilde{\Delta}}{n}  \log\frac{\ell}{\varepsilon}-2 i\pi q}{2\sqrt{\frac{\tilde{\Delta}}{n}  \log\frac{\ell}{\varepsilon}}}\right)+{\rm{Erf}}\left(\frac{\frac{\tilde{\Delta}}{n} \log\frac{\ell}{\varepsilon}+2 i\pi q}{2\sqrt{\frac{\tilde{\Delta}}{n} \log\frac{\ell}{\varepsilon}}}\right) \right)\,,
\label{exact}
\end{eqnarray}
where $\tilde{\Delta}:=g^2/(2\pi)^3$ and Erf$(x)$ is the error function. This formula is however a bit hard to read. It is more common to consider the $\log (\ell/\varepsilon) \gg 1$ expansion instead (or, alternatively, to carry out a saddle point approximation on the integral itself). This approximation was presented  in \cite{GS} for the compactified free boson, where a $U(1)$ symmetry is also present with an associated symmetry field of dimension $(\alpha^2 K)/(8\pi^2)$. Their result, amounts to approximating the sum of error functions by the value $2$ at leading order, giving 
\beq 
\mathcal{Z}_n(q) \approx \frac{1}{2\sqrt{\pi}} \left(\frac{\varepsilon}{\ell}\right)^{4\Delta_n^0} \frac{e^{-\frac{\pi^2 q^2}{|\frac{\tilde{\Delta}}{n}  \log\frac{\varepsilon}{\ell}|}}}{\sqrt{|\frac{\tilde{\Delta}}{n}  \log\frac{\varepsilon}{\ell}|}}\,. \label{squareroot}
\eeq 
Computing now the symmetry resolved R\'enyi entropy as defined by (\ref{sym}), we have that the leading contribution comes from the term $\left(\varepsilon/\ell\right)^{4\Delta_n^0}$, which gives the total R\'enyi entropy of a CFT, i.e. $ (n+1) \log(\ell/\varepsilon)/(6n)$, and the leading correction comes from the square root term in the denominator of  (\ref{squareroot}), giving 
\beq 
S_n(q)\approx \frac{n+1}{6n} \log\frac{\ell}{\varepsilon}-\frac{1}{2}\log \left(4\pi \tilde{\Delta} \log \frac{\ell}{\varepsilon}\right)+\frac{1}{1-n}\log\sqrt{n}\,.
\label{27}
\eeq 
We can also compute the limit $n\rightarrow 1$ to obtain the symmetry resolved von Neumann entropy 
\beq 
S(q)\approx \frac{1}{3} \log\frac{\ell}{\varepsilon}-\frac{1}{2}\log \left(4\pi \tilde{\Delta} \log \frac{\ell}{\varepsilon}\right)-\frac{1}{2}\,.
\label{SRcft}
\eeq 
This formula exhibits all three properties listed at the beginning of this section. We see that the entropy scales logarithmically. Not only that but at leading order, it scales exactly like the total entropy of a CFT, as we know from  \cite{CallanW94, HolzheyLW94, Calabrese:2004eu}. There is however a (negative) double-logarithmic correction, which incorporates some information about the dimension of the symmetry field through the parameter $\tilde{\Delta}$. It has been noted in several places, including in \cite{Horvath:2021rjd} that this double-log term is cancelled out by the number entropy contribution when computing the total entropy through (\ref{number}). Finally, we observe that there is no dependence on the symmetry charge $q$ so that we have {\it equipartition} of entanglement among symmetry sectors. From the derivation, it is clear that this equipartition is not an exact result, but only true in the particular limit of a large region as considered here. Dependence on $q$ is quickly found when considering higher corrections, as we see from the exact formula (\ref{exact})\footnote{The breakdown of equipartition at higher orders has been recently discussed for CFTs with underlying non-invertible symmetries in \cite{Sierra_2024}.}. The dependence on the dimension of the symmetry field has to also be treated with care, since its meaningfulness relies on the assumption that there are no additional non-universal constant corrections to the formula (\ref{SRcft}), or that one knows exactly how to subtract those.  

These results employ explicitly properties of 1+1D CFT, notably the simple scaling of two-point functions of primary fields which is imposed by conformal invariance. Similarly strong constrains exist for three and higher-point functions, though these are much less trivial \cite{DMS}. We may ask how these properties change for a massive QFT, which can be seen as a (massive) perturbation of CFT \cite{Zamolodchikov:1989zs}. The answer is that, in general, away from criticality, it is much harder to find exact formulae for correlation functions, unless we are dealing with free theories. There is however a class of models, namely integrable QFTs (IQFTs), for which there are special methods that we can use to investigate correlation functions of any fields, including CTFs. The leading method is the form factor program \cite{KW,smirnovbook,entropy,Horvath:2020vzs} which we discuss below.

However, we also know since the work \cite{Calabrese:2004eu}, with a later proof in \cite{hastings}, that the entanglement entropy of an interval in a gapped system saturates to a constant value. This value depends on the mass scale/correlation length and this dependence is identical to that found in CFT, with the simple replacement of the interval size $\ell$ by the correlation length $\xi$. This property extends to the symmetry resolved entropies, again in the limit of a very large region, where the dependence on region size drops out. 

There is a very simple way to recover this result, which does not require the use of any specialised techniques, so we will present this derivation before introducing the form factor program. The derivation is based on the property of clustering of correlators. Clustering means the factorisation of multi-point functions into products of vacuum expectation values of fields in local QFT, whenever the distance between fields is very large. In our example it means that the charged moments (\ref{charge}) in a gapped QFT satisfy
\beq 
\lim_{\ell\rightarrow \infty}\varepsilon^{4\Delta_n^\alpha} {}_n\bra 0|\TT_n^\alpha(0)\tTT_n^\alpha(\ell)|0\ket_n = \varepsilon^{4\Delta_n^\alpha}  |{}_n\bra 0|\TT_n^\alpha(0)|0\ket_n|^2\,. 
\eeq 
In 1+1D QFT vacuum expectation values scale as powers of a fundamental mass scale
\beq 
{}_n\bra 0|\TT_n^\alpha(0)|0\ket_n= U_n^\alpha \, m^{2\Delta_n^\alpha}\,,
\eeq 
where $U_n^\alpha$ is a constant\footnote{The value of the VEV is operator and model-dependent but it is also universal in the sense that, once the field and the model are fixed, then its value does not depend on any regulators. If the operator is primary in the UV limit and its CFT correlators are normalised according to conformal normalisation, that is, their two-point function is $\ell^{-4\Delta_n^\alpha}$ as in (\ref{MCFT}), then $U_n^\alpha$ is entirely fixed. See the computation in \cite{entropy} of the value of $U_n^0$ for the Ising field theory.}, so the charged moments scale as
\beq 
Z_n(\alpha)= |U_n^\alpha|^2 \left(m \varepsilon \right)^{4\Delta_n^\alpha}. 
\eeq 
Comparing this last expression with (\ref{MCFT}) we see that the two expressions are identical up to the replacement $\ell \mapsto m^{-1} \, \propto \, \xi$ and the normalisation constant, so that the computation we carried out above can be performed in exactly the same fashion for massive theories and gives the same formulae above with $\log (\ell/ \varepsilon)$ replaced by $\log (m \varepsilon)$ and an additional dependence on a universal constant $U_n^\alpha$.

The conclusion is then that also in massive 1+1D QFT the SREE saturates to a constant value which depends on the correlation length and, for large regions and large correlation length, is independent of the charge, that is, there is again equipartition. The form factor technique mentioned above, allows us to obtain corrections to saturation, which incorporate a dependence on the size of the subsystem, usually leading to exponentially decaying corrections. This kind of analysis has been performed in various papers \cite{Horvath:2020vzs,Horvath:2021rjd,Potts}. Form factor techniques have also been employed in the study of the SREE and negativity of excited states of QFT in \cite{ourPartI,ourPartII,ourPartIII}. We review the main features of this technique in the next section. 

\section{Form Factors and Their Applications}
\label{FFP}
Integrable 1+1D QFTs have many special features, which stem from the interplay between low dimensionality and the presence of infinitely many conservation laws. As a result IQFTs, even interacting ones, are severely constrained. In particular, their scattering amplitudes and the matrix elements of local fields, called form factors, can be obtained exactly. {\it Exactly} means non-perturbatively, as solutions to a set of equations. The program dedicated to systematically solving these equations and testing solutions for consistency is called {\it bootstrap program} and has led to a huge amount of analytical results over many decades, making integrable models some of the most studied theories in mathematical and theoretical physics. For the purposes of this review, we are interested in the form factor equations satisfied by BPTFs and CTFs. The former were first written in \cite{entropy} and the latter in \cite{Horvath:2020vzs}, and both are generalisations of the standard form factor equations for local fields, which are known since the 70s \cite{KW, smirnovbook}. We summarise the main aspects of the program below, largely following the discussion presented in \cite{Horvath:2021rjd}.

The form factors (FF) are matrix elements of (semi-) local operators
${\mathcal O}({\bf x})$ between the vacuum state $|0\ket$ and asymptotic states, here represented by the set of rapidities $\theta_i$ and quantum numbers $\gamma_i$, 
\begin{equation}
F_{\gamma_{1}\ldots\gamma_{k}}^{{\mathcal O}}(\theta_{1},\ldots,\theta_{k}):=\langle0|{\mathcal O}({\bf 0})|\theta_{1},\ldots\theta_{k}\rangle_{\gamma_{1}\ldots\gamma_{k}}.\label{eq:FF}
\end{equation}
In massive field theories like the sine-Gordon model and the Ising field theory introduced earlier, the asymptotic states
are spanned by multi-particle excitations with energy and momentum given by $E_{\gamma}(\theta)= m_{\gamma}\cosh\theta, P_\gamma(\theta)=m_{\gamma}\sinh\theta$,
where $\gamma$ indicates the particle species and $\theta$
its rapidity. In such models, any multi-particle
state can be constructed from the vacuum state $|0\rangle$ as 
\begin{equation}
|\theta_{1},\theta_{2},...,\theta_{k}\rangle_{\gamma_{1}\ldots\gamma_{k}}=A_{\gamma_{1}}^{\dagger}(\theta_{1})A_{\gamma_{2}}^{\dagger}(\theta_{2})\ldots.A_{\gamma_{k}}^{\dagger}(\theta_{k})|0\rangle\:,\label{eq:basis}
\end{equation}
where $A^{\dagger}$s are particle creation operators. In an IQFT with
factorised scattering, that is, where all scattering processes can be factorised into two-particle scattering events, the creation and annihilation operators $A_{\gamma_{i}}^{\dagger}(\theta)$
and $A_{\gamma_{i}}(\theta)$ satisfy the Zamolodchikov-Faddeev (ZF)
algebra \cite{ZA,FA}. In addition, if the theory is non-diagonal, like sine-Gordon, meaning that the incoming and outgoing particles in a two-body scattering process may be distinct, even if they belong to the same mass multiplet, then the algebra is
\begin{eqnarray}
A_{\gamma_{i}}^{\dagger}(\theta_{i})A_{\gamma_{j}}^{\dagger}(\theta_{j}) & = & S_{\gamma_{i}\gamma_{j}}^{\eta_{i}\eta_{j}}(\theta_{i}-\theta_{j})A_{\eta_{j}}^{\dagger}(\theta_{j})A_{\eta_{i}}^{\dagger}(\theta_{i})\:,\nonumber \\
A_{\gamma_{i}}(\theta_{i})A_{\gamma_{j}}(\theta_{j}) & = & S_{\gamma_{i}\gamma_{j}}^{\eta_{i}\eta_{j}}(\theta_{i}-\theta_{j})A_{\eta_{j}}(\theta_{j})A_{\eta_{i}}(\theta_{i})\:,\nonumber \\
A_{\gamma_{i}}(\theta_{i})A_{\gamma_{j}}^{\dagger}(\theta_{j}) & = & S_{\gamma_{i}\gamma_{j}}^{\eta_{i}\eta_{j}}(\theta_{j}-\theta_{i})A_{\gamma_{j}}^{\dagger}(\theta_{j})A_{\gamma_{i}}(\theta_{i})+\delta_{\gamma_{i},\gamma_{j}}2\pi\delta(\theta_{i}-\theta_{j}),\label{eq:ZF}
\end{eqnarray}
where $S_{\gamma_{i}\gamma_{j}}^{\eta_{i}\eta_{j}}(\theta_{i}-\theta_{j})$
denotes the two-body S-matrix of the theory and summation is understood
on repeated indices. The above discussion is general and valid for
any IQFT. In sine-Gordon, 
the particle index $\gamma_{i}$ can take the particular values $s,\bar{s}$ which correspond to the soliton, antisoliton and $b_i$ with $i=1,2 \ldots$ corresponding to their bound states (breathers), on which there will be a finite number determined by the value of the coupling $g$ introduced in (\ref{sgL}). In the Ising field theory on the other hand, there is a single particle type, so the particle index is redundant.

In the $n$-copy (replica) IQFT indices are doubled, in the sense that particles are characterized both by their species ($\gamma_i$, $\eta_i$ in the formulae below) and their copy number ($\mu_i,\nu_i$ in the formulae below). The two-body scattering matrix is then generalised 
to
\begin{equation}
\begin{split}S_{(\gamma_{i},\mu_{i})(\gamma_{j},\mu_{j})}^{(\eta_{i},\nu_{i})(\eta_{j},\nu_{j})}(\theta)= & \delta_{\mu_{i},\nu_{i}}\delta_{\mu_{j},\nu_{j}}\begin{cases}
S_{\gamma_{i}\gamma_{j}}^{\eta_{i}\eta_{j}}(\theta) & \mu_{i}=\mu_{j}\\
\delta_{\gamma_{i},\eta_{i}}\delta_{\gamma_{j},\eta_{j}} & \mu_{i}\neq\mu_{j}
\end{cases}\end{split}
\end{equation}
which defines the replica theory as a set of non-interacting copies of the original model. 

To make our notations easier we introduce
the multi-indices
\begin{equation}
a_{i}=(\gamma_{i},\mu_{i})\quad {\mathrm{with}} \quad \bar{a}_{i}=(\bar{\gamma}_{i},\mu_{i})\quad \mathrm{and}\quad \hat{a}_{i}=(\gamma_{i},{\mu}_{i}+1)\,,
\end{equation}
where $\bar{\gamma}_{i}$ denotes the antiparticle of $\gamma_{i}.$

\subsection{Form Factor Equations for  $U(1)$ CTFs \label{U1CBPTF}}

The main input needed to write the form factor equations for CTFs are the exchange properties of the $U(1)$ CTFs \eqref{U(1)CBPTFSpatialExchange}, similar to how these were used in \cite{entropy} to obtain FF equations for the BPTFs. The result was presented in \cite{Horvath:2021fks, Horvath:2020vzs}. Consider the case when the underlying symmetry is $U(1)$. Then the new FF equations
incorporate a $U(1)$ phase $e^{\frac{i\alpha}{n}}$
in the monodromy properties corresponding the Aharonov-Bohm flux.
Denoting the FFs of $\mathcal{T}_n^{\alpha}({\bf 0})$ by $F^{\alpha}_{a_1\ldots a_k}(\theta_1,\ldots,\theta_k;n)$, the FF equations can be formulated as
\begin{eqnarray}
 &  & F_{\underline{a}}^{\alpha}(\underline{\theta};n)=S_{a_{i}a_{i+1}}^{a'_{i}a'_{i+1}}(\theta_{i\,i+1})F_{\ldots a_{i-1}a'_{i+1}a'_{i}a_{i+2}\ldots}^{\alpha}(\ldots\theta_{i+1},\theta_{i},\ldots ;n),\label{eq:U1FFAxiom1}\\
 &  & F_{\underline{a}}^{\alpha}(\theta_{1}+2\pi i,\theta_{2},\ldots,\theta_{k};n)=e^{\frac{i\kappa_{1}\alpha}{n}}F_{a_{2}a_{3}\ldots a_{k}\hat{a}_{1}}^{\alpha}(\theta_{2},\ldots,\theta_{k},\theta_{1};n),\label{eq:U1FFAxiom2}\\
 &  & -i\underset{\theta_{0}'=\theta_{0}+i\pi}{{\rm Res}}F_{\bar{a}_{0}a_{0}\underline{a}}^{\alpha}(\theta_{0}',\theta_{0},\underline{\theta};n)=F_{\underline{a}}^{\alpha}(\underline{\theta};n),\label{eq:U1FFAxiom3}\\
 &  & -i\underset{\theta_{0}'=\theta_{0}+i\pi}{{\rm Res}}F_{\bar{a}_{0}\hat{a}_{0}\underline{a}}^{\alpha}(\theta_{0}',\theta_{0},\underline{\theta};n)=-e^{\frac{i\kappa_{0}\alpha}{n}}\mathcal{\mathcal{S}}_{\hat{a}_{0}\underline{a}}^{\underline{a'}}(\theta_{0},\underline{\theta},k)F_{\underline{a'}}^{\alpha}(\underline{\theta};n),\nonumber \\
 &  & -i\underset{\theta_{0}'=\theta_{0}+i\bar{u}_{\gamma\delta}^{\varepsilon}}{{\rm Res}}F_{(\gamma,\mu_{0})(\delta,\mu'_{\text{0}})\underline{a}}^{\alpha}(\theta_{0}',\theta_{0},\underline{\theta};n)=\delta_{\mu_{0},\mu'_{0}}\Gamma_{\gamma\delta}^{\varepsilon}F_{(\varepsilon,\mu_{0})\underline{a}}^{\alpha}(\theta_{0},\underline{\theta};n),\label{eq:U1FFAxiom4}
\end{eqnarray}
where several short-hand notations have been used: as usual $\theta_{ij}=\theta_{i}-\theta_{j}$,
$\underline{\theta}:= \theta_{1},\theta_{2},...,\theta_{k}$
and $\underline{a}:=(\gamma_{1},\mu_{1})(\gamma_{2},\mu_{2})\ldots(\gamma_{k},\mu_{k})$. The factor in the fourth equation is an abbreviation for
\begin{equation}
\mathcal{\mathcal{S}}_{\hat{a}_{0}\underline{a}}^{\underline{a'}}(\theta_{0},\underline{\theta},k)=S_{{a}_{0}a_{1}}^{c_{1}a'_{1}}(\theta_{01})S_{c_{1}a_{2}}^{c_{2}a'_{2}}(\theta_{02})\ldots S_{c_{k-1}a_{k}}^{{a}_{0}a'_{k}}(\theta_{0k})\,.\label{eq:CalP}
\end{equation}
Equation (\ref{eq:U1FFAxiom1}) is the {\it exchange relation} while (\ref{eq:U1FFAxiom2}) is the {\it crossing relation}. Together they form what is known as {\it Watson's equations}, which constraint the monodromy of the FFs. Equation (\ref{eq:U1FFAxiom3}) and the equation below it are {\it kinematic residue equations} while (\ref{eq:U1FFAxiom4}) is the {\it bound state residue equation}. Together, they determine the pole structure of the FFs as well as leading to recursive discrete equations on the particle number. Equation (\ref{eq:U1FFAxiom4}) involves the parameters $u_{\gamma \delta}^\varepsilon$ and $\Gamma_{\gamma \delta}^\varepsilon$. For the purposes of this review, it is not essential to define those in detail. Suffice it to say that they are parameters relating to the position of the bound state poles of the $S$-matrix and to value of the associated residue. See \cite{Horvath:2021rjd} for further details and also equation (\ref{48}) and the paragraph thereafter. 

The CTF is generally spinless and therefore, by Lorenz invariance, the form factors are functions of rapidity differences only. It means that the variable dependence of the FFs can be reduced by one. The index $\kappa$ in the phase factors
corresponds the $U(1)$ charge of the corresponding particle. In the sine-Gordon model,  the $\kappa$ index takes three possible values
\begin{equation}
\begin{split}\kappa_{i}= & \begin{cases}
1, & \quad \gamma_{i}=s,\\
-1, & \quad \gamma_{i}=\bar{s},\\
0, & \quad \gamma_{i}=b_{j},
\end{cases}\end{split}
\end{equation}
which means that the non-trivial monodromy does not affect the breather
sector of the theory. Note also that for $\alpha=0$ we recover the equations for the standard BPTF, whose form factors in the sine-Gordon case were studied in \cite{Castro-Alvaredo:2008usl,Castro-Alvaredo:2021jjl},  and for $n=1$ we have instead the FF equations of the sine-Gordon $U(1)$ field, whose solutions have also been studied \cite{KK,Luky1,Takacs:2008zz}.

From this point, the solution procedure is standard in the context of the FF program. We start by obtaining lower-particle FFs, and focusing on a single copy and then use the FF equations recursively to access higher particle numbers and other replica numbers. The one-particle
FFs when non-vanishing, are rapidity independent and the two-particle
ones depend only on the rapidity difference. Akin to the BPTF, the novel
composite field is neutral in relation to the sine-Gordon $U(1)$-symmetry,
which implies the vanishing of any FFs involving
a different number of solitons and antisolitons. In particular one
finds that 
\begin{equation}
F_{ss}^{\alpha}(\theta;n)=F_{\bar{s}\bar{s}}^{\alpha}(\theta;n)=F_{\bar{s}b_{k}}^{\alpha}(\theta;n)=F_{{s}b_{k}}^{\alpha}(\theta; n)=F_{s}^{^{\alpha}}(n)=F_{\bar{s}}^{\alpha}(n)=0\,,\quad\forall\quad k\in\mathbb{Z}^{+}\,.
\end{equation}
There are however non-vanishing one-particle and two-particle FFs for all breather combinations. 

Under these considerations, Watson's equations \eqref{eq:U1FFAxiom1},
\eqref{eq:U1FFAxiom2} for non-vanishing two-particle form factors
and particles in the same copy can be summarised as 
\begin{eqnarray}
F_{s\bar{s}}^{\alpha}(\theta;n) & = & S_{s\bar{s}}^{s\bar{s}}(\theta)F_{\bar{s}s}^{\alpha}(-\theta;n)+S_{s\bar{s}}^{\bar{s}s}(\theta)F_{s\bar{s}}^{\alpha}(-\theta; n)=e^{i\alpha}F_{\bar{s}s}^{\alpha}(2\pi in-\theta; n),\label{fpm1U1}\\
F_{\bar{s}s}^{\alpha}(\theta; n) & = & S_{\bar{s}s}^{\bar{s}s}(\theta)F_{s\bar{s}}^{\alpha}(-\theta;n)+S_{\bar{s}s}^{s\bar{s}}(\theta)F_{\bar{s}s}^{\alpha}(-\theta; n)=e^{-i\alpha}F_{s\bar{s}}^{\alpha}(2\pi in-\theta; n),\label{fmp1U1}\\
F_{b_{i}b_{j}}^{\alpha}(\theta; n) & = & S_{b_{i}b_{j}}(\theta)F_{b_{i}b_{j}}^{\alpha}(-\theta; n)=F_{b_{i}b_{j}}^{\alpha}(2\pi in-\theta; n)\quad\mathrm{for}\quad{i-j}\in2\mathbb{Z}\,.\label{fpm2U1}
\end{eqnarray}
The last equality in the top two equations requires the use of the crossing property (the second of Watson's equations) $n$ times, leading to the phases $e^{\pm i\alpha}$. These two sets of equations can be solved by diagonalisation of the soliton-antisoliton sector, as done already for the BPTF in \cite{Castro-Alvaredo:2008usl}. 

The kinematic
residue equations \eqref{eq:U1FFAxiom3} are 
\begin{equation}
-i\underset{\theta=i\pi}{{\rm Res}}F_{s\bar{s}}^{\alpha}(\theta; n)=-i\underset{\theta=i\pi}{{\rm Res}}F_{b_{i}b_{i}}^{\alpha}(\theta; n)={}_n\langle 0|\mathcal{T}_n^{\alpha}|0\rangle_n\,\quad\forall\quad i\in\mathbb{N}\,.\label{kinU1}
\end{equation}
Finally,
the bound state residue equations \eqref{eq:U1FFAxiom4} are 
\begin{equation}
-i\underset{\theta=i\pi u_{s\bar{s}}^{c}}{{\rm Res}}F_{s\bar{s}}^{\alpha}(\theta; n)=\Gamma_{s\bar{s}}^{c}F_{c}^{\alpha}(n),\label{DynPoleU1}
\end{equation}
where $c$ is any particle that is formed as a bound state of $s+\bar{s}$
for rapidity difference $\theta=i\pi u_{s\bar{s}}^{c}$. For the breather
sector it is again convenient to write the more general equation 
\begin{equation}
-i\underset{\theta=\theta_{0}}{{\rm Res}}F_{b_{i},b_{j},\underline{a}}^{\alpha}(\theta+iu,\theta_{0}-i\tilde{u},\underline{\theta};n)=\Gamma_{b_{i}b_{j}}^{b_{i+j}}F_{b_{i+j},\underline{a}}^{\alpha}(\theta,\underline{\theta}; n)\,,
\label{48}
\end{equation}
where $\underline{a}$ is any particle combination for which the FF is non-vanishing. We recall that $u+\tilde{u}=u_{ij}^{i+j}$
and $\theta=i\pi u_{ij}^{i+j}$ is the pole of the scattering matrix
$S_{b_{i}b_{j}}(\theta)$ corresponding to the 
formation of breather $b_{i+j}$, and $u$ and $\tilde{u}$ are related to
the poles of $S_{b_{j}b_{i+j}}(\theta)$ and $S_{b_{i}b_{i+j}}(\theta)$, respectively.
It is important to emphasise that the bootstrap equations \eqref{eq:U1FFAxiom1}-\eqref{eq:U1FFAxiom4}
or \eqref{fpm2U1}-\eqref{DynPoleU1} for the $U(1)$ neutral breathers
are identical to those of the conventional BPTFs, nevertheless the
FFs are different from those of $\mathcal{T}$.

Finally, two-particle FFs with arbitrary replica
indices can be easily obtained from the above  through relations which are themselves a consequence of Watson's equations. They can be written as
\begin{equation}
F_{(\gamma,j)(\eta,k)}^{\alpha}(\theta; n)=\begin{cases}
e^{-i\alpha\kappa_{\eta}/n}F_{\eta\gamma}^{\alpha}(2\pi i(k-j)-\theta; n) & \text{ if }k>j,\\
e^{i\alpha\kappa_{\gamma}/n}F_{\gamma\eta}^{\alpha}(2\pi i(j-k)+\theta; n) & \text{otherwise,}
\end{cases}\label{eq:FD2FullU1}
\end{equation}
where $\kappa_{\gamma}$ is zero for neutral breathers. The two-particle FFs of
the other field $\tilde{\mathcal{T}}_n^{\alpha}$ denoted by $\tilde{F}^{\alpha}_{a_1a_2}(\theta;n)$
can be simply written as
\begin{equation}
\tilde{F}_{(\gamma,j)(\eta,k)}^{\alpha}(\theta;n)=F_{(\gamma,n-j)(\eta,n-k)}^{-\alpha}(\theta;n)\,.
\end{equation}
Solutions for these form factors were reported in \cite{Horvath:2021rjd}. Even for one and two particles they are rather involved, given as they are in terms of integral representations, so we will not reproduce them here, as they are beyond the scope of this review. 

\subsection{Form Factor Equations for $\mathbb{Z}_2$ CTFs \label{Z2CBPTF}}
Although the equations (\ref{eq:U1FFAxiom1})-(\ref{eq:U1FFAxiom4}) are rather general, they are also specific for $U(1)$ symmetry, which is a continuous symmetry. There are however many integrable models possessing discrete symmetries such as $\mathbb{Z}_k$ for some integer $k$. For example, the family of Potts models includes models with different values of $k$. The simplest non-trivial case is $k=3$, which was studied in \cite{Potts}. Another well-known case is the Ising field theory, introduced in subsection \ref{Ising}. The Ising model has $\mathbb{Z}_2$ symmetry and the FF equations for the CTFs $\TT_n^\pm$ introduced earlier are much simpler than (\ref{eq:U1FFAxiom1})-(\ref{eq:U1FFAxiom4}). Not only is the ``symmetry phase" just $\pm 1$ but the $S$-matrix itself is also $\pm 1$ depending on whether particles are in the same or distinct copies. For the field $\TT_n^-$ the equations where written in \cite{Horvath:2020vzs} but they can be written for both fields as
\begin{eqnarray}
 &  & F_{\underline{a}}^{\pm}(\underline{\theta};n)=S_{a_{i}a_{i+1}}(\theta_{i\,i+1})F_{\ldots a_{i-1}a'_{i+1}a'_{i}a_{i+2}\ldots}^{\pm}(\ldots\theta_{i+1},\theta_{i},\ldots ;n),\label{eq:UIFFAxiom1}\\
 &  & F_{\underline{a}}^{\pm}(\theta_{1}+2\pi i,\theta_{2},\ldots,\theta_{k};n)=\pm F_{a_{2}a_{3}\ldots a_{k}\hat{a}_{1}}^{\pm}(\theta_{2},\ldots,\theta_{k},\theta_{1};n),\label{eq:UIFFAxiom2}\\
 &  & -i\underset{\theta_{0}'=\theta_{0}+i\pi}{{\rm Res}}F_{\bar{a}_{0}a_{0}\underline{a}}^{\pm}(\theta_{0}',\theta_{0},\underline{\theta};n)=F_{\underline{a}}^{
 \pm}(\underline{\theta};n),\label{eq:UIFFAxiom3}\\
 &  & -i\underset{\theta_{0}'=\theta_{0}+i\pi}{{\rm Res}}F_{\bar{a}_{0}\hat{a}_{0}\underline{a}}^{\pm}(\theta_{0}',\theta_{0},\underline{\theta};n)=\mp F_{\underline{a'}}^{\pm}(\underline{\theta};n)\,.
\end{eqnarray}
Here, the particle indices are just copy numbers since there is a single particle type, so the $S$-matrix is simply
\beq 
S_{ab}(\theta)=(-1)^{\delta_{ab}}\,.
\eeq 
For the $+$ sign these are the same equations as for the BPTF $\TT$ of the Ising model, whose FFs were studied in various papers \cite{entropy,nexttonext}. However, since the field $\TT^+_n$ still has different conformal dimension from $\TT$, its FFs are distinct from those of $\TT$. They were briefly discussed in \cite{CTFMichele}. The FFs of $\TT^-_n$ were obtained in \cite{Horvath:2020vzs}, and they have interesting summation properties which were discussed in \cite{CTFMichele}. Since they are relatively simple functions, we give here the formula for the two-particle form factor of two fermions living in the same copy
\beq 
F_{11}^{-}(\theta; n)= 
\frac{{}_n \bra 0| \TT_n^-({\bf 0})|0 \ket_n\sin\frac{\pi}{n}}{2n \sinh\left(\frac{i\pi - \theta}{2n}\right)\sinh\left(\frac{i\pi + \theta}{2n}\right)}\frac{\sinh\frac{\theta}{n}}{\sinh\frac{i\pi}{n}}\,.
\label{wfun}
\eeq
We note how the denominator produces the two expected kinematic poles at the values of rapidities $\theta=i\pi$ and $\theta=i\pi(2n-1)$, while the function $\sinh (\theta/n)$ ensures that the FF is an odd function of $\theta$ (as required by the exchange property) and picks up a minus sign under crossing. 

\section{Finite Interval Corrections}
\label{corrections}
The computation of corrections to the ``saturation" formulae that are obtained by replacing $\ell\mapsto m^{-1}$ in (\ref{27})-(\ref{SRcft}) is a mathematically very technical exercise, which employs the full power of the FF program and the solutions to the equations presented in the previous section. However, the main ideas are relatively simple and can be summarised here. We refer the reader to the books \cite{smirnovbook,mussardobook} for a more detailed review of the FF program and its applications to computing correlators.  In the ground state of a massive IQFT with a single-particle spectrum, two-point functions may be spanned in terms of the absolute values of the FFs above. For a local, spinless field $\mathcal{O}({\bf x})$ the first few terms of such an expansion look like
\beqa 
\bra 0|\mathcal{O}({\bf 0})\mathcal{O}^\dagger({\bf x})|0\ket &=& |\bra 0|\mathcal{O}({\bf 0})|0\ket|^2 + \int_{-\infty}^\infty \frac{d\theta}{2\pi} |F_1|^2 e^{- m\ell \cosh\theta} \nonumber\\
&& + \frac{1}{2}\int_{-\infty}^\infty \frac{d\theta_1 d\theta_2}{(2\pi)^2} |F_2(\theta_1-\theta_2)|^2 e^{-m\ell \cosh\theta_1 -m\ell  
\cosh \theta_2} + \cdots
\eeqa 
where $m$ is the mass of the particle, ${\bf x}:=(0,-i\ell)$ is the space-time position, $F_1$ is the (rapidity-independent) one-particle FF, $F_2(\theta)$ is the two-particle FF etc. Higher corrections, will include higher-particle form factors. The expansion is simply obtained by ``inserting" a sum over a complete set of states between the two fields in the correlator. This expansion is usually described as an ``infrared" expansion, meaning that it is rapidly convergent for large $\ell m$. 

Depending on the symmetries present in the theory, it is possible that some of the FFs are vanishing. For example, in the Ising field theory, the fermion field has a single non-vanishing FF for one particle, whereas the ``energy" field (the mass term in the action (\ref{AIsing})), which is bilinear in the fermions has only a two-particle FF. On the other hand, the twist fields $\sigma$ and $\mu$ introduced earlier have non-vanishing odd- or even-particle FFs, as a result of $\mathbb{Z}_2$ symmetry. 

When the one-particle FF is non-vanishing, we see directly that the first $\ell$-dependent contribution is a Bessel function $K_0(m\ell)$, which is obtained by performing the single integral above
\beq 
\int_{-\infty}^\infty \frac{d\theta}{2\pi} |F_1|^2 e^{-r m \cosh\theta} =\frac{|F_1|^2}{\pi} K_0(m\ell)\,,
\eeq 
leading to exponentially decaying corrections for large $m\ell$. 
Should $F_1=0$ then the first correction will come from the two-particle FF. Since the FF only depends on rapidity differences, it is possible to change variables to $\theta=\theta_1-\theta_2$ and $\hat{\theta}=\theta_1+\theta_2$ and integrate out $\hat\theta$ leaving a contribution
\beq 
\int_{-\infty}^\infty \frac{d\theta}{(2\pi)^2} |F_2(\theta)|^2 K_0\left(2 m\ell \cosh\frac{\theta}{2}\right)\,. 
\eeq 
For large $\ell m$ this contribution is also exponentially decaying, as are higher particle ones.  

Although the case of BPTFs and CTFs is considerably more complicated than this, the generalisation is straightforward. Let us rewrite the formula above for the field $\TT_n^-$ we discussed in the previous section for the Ising field theory
\beqa 
\bra 0|\TT_n^-({\bf 0})\tTT_n^-({\bf x})|0\ket =|\bra 0|\TT_n^-({\bf 0})|0\ket|^2  + \sum_{j=1}^n \sum_{k=1}^n\int_{-\infty}^\infty \frac{d\theta}{(2\pi)^2} |F^-_{jk}(\theta;n)|^2 K_0\left(2 m\ell \cosh\frac{\theta}{2}\right)+ \cdots
\eeqa 
The main change is that we now have sums over copy numbers (in the sine-Gordon model we would also have sums over particle types). For this field, we also have a vanishing one-particle FF so the first non-trivial correction will come from the two-particle FF. A number of simplifications are possible: first, all copies are identical so at least one of the sums can be eliminated and replaced by an overall factor $n$. Then, the form factor $F_{jk}^-(\theta;n)$ can be related to the FF $F_{11}^-(\theta;n)$ by employing identities akin to (\ref{eq:FD2FullU1}). Calling 
\beq 
C^-(\ell;n):= \frac{\bra 0|\TT_n^-({\bf 0})\tTT_n^-({\bf x})|0\ket -|\bra 0|\TT_n^-({\bf 0})|0\ket|^2}{|\bra 0|\TT_n^-({\bf 0})|0\ket|^2}\,,
\eeq 
the normalised connected correlator, we find that, at two-particle order, it can be written as
\beq 
C^-(\ell;n):=n|\bra 0|\TT_n^-({\bf 0})|0\ket|^{-2}\sum_{j=0}^{n-1} \int_{-\infty}^\infty \frac{ \mathrm{d}\theta_1 \mathrm{d}\theta_2}{(2\pi)^2} |F^-_{11}(-\theta+2\pi i j;n)|^2  K_0\left(2 m\ell \cosh\frac{\theta}{2}\right)\,.
\label{crr}
\eeq 
This expression can be easily evaluated numerically, with the FF (\ref{wfun}). This evaluation, combined with the evaluation of the same correlator for the BPTF, allows us to compute the SREEs of the two symmetry sectors. The calculation is presented in detail in Chapter 6 of \cite{Horvath:2020vzs}. 

An interesting feature of this type of calculation in the context of entanglement measures is that in order to computer the (symmetry resolved) entanglement entropies we need to take the limit $n\rightarrow 1$ of all the formulae above. This is non-trivial, given that the whole twist field construction is based around the notion of an integer number of replicas $n$. In order to perform the limit $n\rightarrow 1$, expressions such as (\ref{crr}) need to be analytically continued first to $n$ positive and real. A suitable way to do this was presented in \cite{entropy} for the BPTF and applies to CTFs in a similar manner. The idea is to replace the sum in (\ref{crr}) by a contour integral whose value reproduces the sum, as a sum over residues, plus an additional contribution resulting from the kinematic poles of the FF. A beautiful feature of this approach is that once the analytic continuation has been performed, the limit $n\rightarrow 1$, gives rise to formulae that contain a $\delta(\theta)$ term. Thanks to this term the integral (\ref{crr}) (or part of it) can be computed exactly and the exponential corrections obtained in a very simple and universal form. If we call $C^0(\ell;n)$ the connected correlator of the BPTF, the procedure sketched above led to the result
\beq 
-\lim_{n\rightarrow 1}\frac{\partial}{\partial n}C^0(\ell;n)= -\frac{1}{8} K_0(2m \ell) +\mathcal{O}(e^{-3m\ell})
\eeq 
whereas, for the normalised connected correlator $C^-(r;n)$ a similar computation was carried out in \cite{Horvath:2020vzs} giving instead 
\beq 
-\lim_{n\rightarrow 1}\frac{\partial}{\partial n}C^-(\ell;n)= -\frac{1}{8} K_0(2m \ell) + \frac{1}{4\pi} \frac{e^{-2m\ell}}{m\ell}+\mathcal{O}((m\ell)^{-2} e^{-2m\ell})\,.
\eeq 
As we can see, there is one common term proportional to $K_0(2m\ell)$ in both cases. This term is highly universal as it results from the pole structure of the FF and not specific details of the model. For the BPTF the same formula is found for any 1+1D QFT (even non-integrable ones) if $m$ is the mass of the lightest excitation(s) in the theory \cite{next}\footnote{For generic theories, there is a sum of Bessel functions for the different masses, but for large $m\ell$ the lightest particle still provides the leading contribution}.
These formulae can then be employed to obtain $\ell$-dependent corrections to the saturation formulae. These kinds of computations have been carried out in \cite{Horvath:2020vzs} for the Ising and sinh-Gordon models, both exhibiting $\mathbb{Z}_2$ symmetry, for the Potts model in \cite{Potts} where $\mathbb{Z}_3$ symmetry was considered, for the sine-Gordon model \cite{Horvath:2021rjd} and for free theories with $U(1)$ symmetry in \cite{Horvath:2021fks}.

\section{Beyond the Ground State}
\label{excitedstates}
So far, we have focused on ground states. In this section, we want to generalise this study to excited states. Treating these in complete generality would be very difficult since we expect the SREE to depend on many features of the excitations, including their energy, momentum and quantum numbers. However, in special cases, there are interesting universal features associated with the presence of a finite number of excitations within a infinite quantum system. 

Let us start by reviewing some of the existing literature. There are by now many works concerning excited states in low-energy CFT \cite{german1, german2}, critical systems \cite{Vincenzo2}, gapped quantum spin chains \cite{Vincenzo}, and free IQFTs \cite{excited, excited1, excited2, excited3, Zhang1, Zhang_2021a,Zhang2,Zhang3,Zhang4,Mussardo:2021gws}. All of these works deal with standard (ie. non symmetry resolved) measures, nonetheless some useful conclusions can be drawn. On the one hand, the entanglement entropy of excited states with an infinite number of particles in QFT or quantum lattices (i.e. for finite-density excited states) is dominated by the thermodynamic entropy of the corresponding Gibbs state, and satisfies a volume law \cite{EERev2}. On the other hand, this no longer applies to excited states described by a finite number of excitations. In that case, what we have instead is a finite ``excess entropy", that is, the entropy is increased by a finite amount w.r.t. the ground state. This finite amount takes a universal and very simple form, provided that the ratio of total system size to subsystem size is kept finite, while the individual sizes are very large. In this limit, the excess entropy depends only on the relative size of the regions and the particle statistics. In \cite{excited,excited1} this was proven for free QFTs but there is evidence, including in those references, that it is much more general, extending also to interacting and higher dimensional models. As argued in \cite{Mussardo:2021gws}, this is related to the fact that the excess entropy contributions are in essence of a semi-classical nature, therefore highly universal. 

Let us consider how these properties generalise to the symmetry-resolved context. Consider again a bipartite system of total length $L$ and a connected region of length $\ell$. Consider also the scaling limit in which both lengths tend to infinity while keeping the ratio $r=\ell/L$ constant. The system is in a pure state formed of a finite number of excitations, i.e. a zero-density state. The ratio of the charged moments of an excited state $|\Psi\ket$ and those of the ground state $|0\ket$ of the theory can be written as
\beq
\lim_{L \rightarrow \infty} \frac{{}_n\bra \Psi|\TT_n^\alpha(0)\tTT_n^\alpha(rL)|\Psi\ket_n}{{}_n\bra 0|\TT_n^\alpha(0)\tTT_n^\alpha(rL)|0\ket_n}=:M_n^\Psi(r;\alpha)\,,
\label{eq_Mratio}
\eeq
and, as the equation suggests, in the infinite volume limit, it is a function of $r, \alpha, n$ and the state. However, the only state features that play a role in this limit is the number of excitations and their statistics, as we will see below. 

Here we will consider massive IQFTs  although similar results have been obtained in CFT \cite{Capizzi_2020}.
In the former case, whether or not symmetries are present, there arise some technical difficulties when computing the type of correlators involved in \eqref{eq_Mratio}. This is so because whenever one introduces a set of intermediate states to express the correlator as a product of form factors, $\delta$-function singularities appear if the momenta of the intermediate particles coincide with those of the particles in the excited state. To avoid this problem, finite volume form factors must be calculated first, and then the volume must be used as a regulator. But, although a finite volume form factor program for generic local fields exists \cite{PT1,PT2}, its extension to twist fields is still an open problem. Alternatively, a solution was found in \cite{excited2} for complex free theories. It is based on the fact that in that case one can diagonalise the permutation symmetry in the $n$-copy theory and express the BPTF as a product of simpler $U(1)$ twist fields, whose form factors are well-known in the literature \cite{doubling, smirnovbook, YZam}. In finite volume the particle momenta are quantised according to the Bethe-Yang equations, and this plays a crucial role in computations. Using all of these ideas,  the ratio of charged moments (\ref{eq_Mratio}) has been computed in \cite{ourPartI}.  For $U(1)$ symmetry, the formula depends as well as on the parameter $\alpha$, as implicit in (\ref{eq_Mratio}). The results can be summarised as follows:
\begin{enumerate}
    \item For states of $k$ identical (bosonic) excitations of charge $\epsilon$:
\beq
M_n^{k^\epsilon}(r;\alpha)= \sum_{j=0}^k [f_j^k(r)]^n e^{2\pi i j \epsilon \alpha}\,,
\label{for1}
\eeq
where $ f_j^k(r):={}_kC_{j} \, r^j (1-r)^{k-j}$ and ${}_kC_{j}=k!/[j!(k-j)!]$ is the binomial coefficient. Notice that $\epsilon=\pm$ represent the $U(1)$ charge of the bosons. For generic states comprising $s$ groups of $k_i^{\epsilon_i}$ identical particles of charge $\epsilon_i$ we will have 
\beq
M_n^{k_1^{\epsilon_1}\ldots k_s^{\epsilon_s}}(r;\alpha)=\prod_{i=1}^s M_n^{k_i^{\epsilon_i}}(r;\alpha)\,.
\label{general}
\eeq
\item For states containing $k$ distinct excitations either bosonic or fermionic we have instead:
\beqa
M^{1^{\epsilon_{1}}  \dots 1^{\epsilon_k}}_n(r;\alpha)=\prod_{j=1}^k\left[e^{2\pi i \epsilon_j\alpha}r^n+(1-r)^n\right]\,.
\label{re2}
\eeqa 
\end{enumerate}
These results can be derived rather easily if one considers the qubit ``picture" for multiparticle excited states, first employed in \cite{excited}. In this picture, the contributions from the excited state are encoded into a “multi-qubit state” which is built as a combination of states labelled by integers $j$ where we identify the qubit state $j$ with the presence of $j$ identical particles in the entanglement region $A$ i.e.:
\begin{equation}
|\Psi_{\textrm{qb}}\ket = \sum_{\textbf{\textrm{q}} \in \prod_{j \geq 1} \{0,1,\dots,j\}^{N_j}} \sqrt{\prod_{i} f_{j_i}^{q_i}(r)}\,\, |\textbf{\textrm{q}}\ket \otimes |\bar{\textbf{\textrm{q}}}\ket.
\label{qubits}
\end{equation}
The coefficients in the above expansion  represent the probability of finding a specific particle configuration $\textbf{\textrm{q}}=\{q_i: i=1,\ldots,N\}$ in the corresponding entanglement region, and are given by binomial coefficients (see \eqref{for1}), which underlie the (un)distinguishability of excitations\footnote{Considering the bipartite Hilbert space $\mathcal{H}= \mathcal{H}_A \otimes \mathcal{H}_{\bar{A}}$, where each factor can be related to the Hilbert space for $N_j$ sets of $j$ indistinguishable qubits (with $N=\sum_j N_j$).}. Notice that $|\bar{\textbf{\textrm{q}}}\ket$ is the state where the qubits are inverted. For example, the qubit state
\beq 
|1\ket := \sqrt{r} |10\ket + \sqrt{1-r} |01\ket\,\label{1}
\eeq
can be seen as representing a single excitation which is localised in region $A$ with probability $r$ and in region $\bar{A}$ with probability $1-r$.
Moreover, if the charge operator associated with the internal symmetry is $Q=Q_A \otimes 1_{\bar{A}}+1_A \otimes Q_{\bar{A}}$, then 
\begin{eqnarray}
e^{2\pi i \alpha Q_A} |\Psi_{\textrm{qb}}\ket &=& e^{2\pi i \alpha Q_A} \sum_{\textbf{\textrm{q}}} \sqrt{p_{\bf q}}|\textbf{\textrm{q}}\ket \otimes |\bar{\textbf{\textrm{q}}}\ket = \sum_{\textbf{\textrm{q}}} e^{2\pi i \alpha (n^{+}_{q}-n^{-}_{q})} \sqrt{p_{\bf q}} |\textbf{\textrm{q}}\ket \otimes |\bar{\textbf{\textrm{q}}}\ket\,,
\end{eqnarray}
where the summation is over the same set  $\textbf{\textrm{q}}$ as above and $n^\pm_q$ is the number of positively/ negatively charged particles in subsystem $A$  for a particular configuration ${\bf q}$ with probability $p_{\bf q} = \prod_i f_{j_i}^{q_i}(r)$. Consequently, the entropy of this qubit state is the Fourier transform of
\begin{equation}\label{qubitS}
\mathrm{Tr}(\rho_\textrm{q}^n e^{2\pi i\alpha Q_A}) =  \sum_{\bar{\textbf{\textrm{q}}}^\prime}\sum_{\textbf{\textrm{q}}} e^{2\pi i\alpha (n^+_{\textrm{q}}-n^-_{\textrm{q}})} \, p_{\textbf{\textrm{q}}}^n\,   \delta_{\bar{\textrm{q}},\bar{\textrm{q}}^\prime}\,,
\end{equation}
which for the state (\ref{1}) gives (\ref{for1}) with $k=1$.
In this way, in the scaling limit the charged moments ratios (\ref{for1}) and (\ref{general}) equate the charged moments of simple qubit states \eqref{qubitS}.

Once the ratio (\ref{eq_Mratio}) is known,  the SREE of the excited state is obtained from the Fourier transform  of
\beq
Z^{\Psi}_n(r;\alpha)=Z^{0}_n(\alpha) M_n^\Psi(r;\alpha)\, ,
\eeq  
with $Z^{0}_n(\alpha):=\varepsilon^{4\Delta_\alpha} \bra \TT_n^\alpha \ket^2\,$ for 1+1D IQFTs, as introduced earlier. Notice that the ground state correlator will vary depending on the theory considered. But since for free theories $M_n^\Psi(r;\alpha)$ depends on $\alpha$ in a simple manner, it is possible to express the SREE of the excited state fully in terms of the ground state (see \cite{ourPartI} for further details). Notice that, in the qubit picture (as opposed to QFT) the ground state is trivial and the results \eqref{qubitS} are directly the charged moments of the excited state. 


To conclude, we would like to note that in \cite{ourPartII} numerical evidence for the validity of the formulae in this section has been provided for a 1D Fermi gas and a complex harmonic chain. Interestingly, the former example is a critical theory, suggesting that these results also hold for massless/gapless models, as shown in \cite{Capizzi_2020} for the total entropy. The results have been extended to interacting models (magnon states) and higher-dimensional theories \cite{ourPartII, excited3} and other measures, such as the logarithmic negativity \cite{ourPartIII}. In general, the same formulae are found whenever excitations are localised, in the sense that their characteristic length, be it the correlation length or the De Broglie wavelength, are small compared to subsystem size. 


\section{Conclusion}
\label{conclu}
In this review article we have presented a summary of some of the main results concerning the symmetry resolved entanglement entropy of 1+1D QFTs. This is an entanglement measure, that quantifies the contribution to the total entanglement entropy that is due to separate symmetry sectors, provided the theory has an internal symmetry. We have revisited the key definitions and techniques, and presented a literature review. 

As we have seen, the relationship between entanglement measures and correlation functions of symmetry fields plays a major role. Through this relationship it is possible to easily derive the main properties of the SREE, such as its equipartition in CFT and its saturation in massive QFT. More detailed information about specific models, system-size dependence etc. can be gathered by employing the form factor program for twist fields. 

The study of symmetry resolved entanglement measures has put the focus on symmetries and their role in the context of entanglement. This has given rise not only to an enormous amount of publications, containing results of analytical, numerical and even experimental nature, but also to a renewed interest in the role of symmetries, their breaking and how entanglement measures can help us capture information about both. In connection to symmetry breaking, the concept of {\it entanglement asymmetry} has lately emerged \cite{Ares:2022koq}. Here the interest lies in quantifying the extend to which a symmetry is broken and the speed with which it is restored by studying the evolution of entanglement, particularly in the wake of a symmetry-breaking quantum quench. Several interesting studies of this quantity have been carried out in the past two years \cite{Capizzi:2020mio,Ares:2023kcz,Bertini:2023ysg,Capizzi:2023xaf,Ares:2023ggj}.

Length constrains have kept us from discussing many interesting results and approaches, one clear omission being the other symmetry resolved entanglement measures. We hope nonetheless that our review will be useful to those wanting to learn some introductory facts about this popular area of research. 
\medskip

\noindent {\bf Acknowledgements:} 
 We would like to thank K.K. Phua
(Chairman and Editor-in-Chief) and Rucha Shaha (Journal Publishing Editor)
from World Scientific Publishing, for their invitation to submit this brief review to Modern Physics Letters B and their support throughout the writing process. 

The opportunity to write this review came about thanks to Olalla's participation at last year's APS March Meeting (Las Vegas, March 2023).
Olalla would like to thank the organisers of the meeting for the invitation to present work on symmetry resolved measures at a session specially dedicated to the topic. Special thanks to the organiser of that session, Prof. Adrian Del Maestro, for bringing together an international community of experts on this specialised field of mathematical physics. 

Olalla and Luc\'ia have collaborated on the papers \cite{ourPartI, ourPartII} on symmetry resolved entanglement measures of excited states of quantum field theory and some of those results are reviewed in Section \ref{excitedstates}. These were completed with collaborators Luca Capizzi, Cecilia De Fazio and Michele Mazzoni, and heavily inspired by previous work with Benjamin Doyon and Istv\'an M. Sz\'ecs\'enyi \cite{excited,excited1,excited2,excited3}. We thank them all for their various collaborations.  We would also like to thank Moshe Goldstein and Eran Sela for their insightful explanation of one aspect of their work \cite{GS}.

Olalla is grateful to EPSRC for financial support under a Small Grant EP/W007045/1.
Lucía is grateful to the Regional Government of Castilla y León (Junta de Castilla y León), and the Spanish
Ministry of Science and Innovation MICIN for the financial support received through the ``Q-CAYLE:
Secure Quantum Communications in Castilla y León" project, also co-funded by the European Union NextGenerationEU (PRTR C17.I1).

Olalla has worked on entanglement measures in quantum field theory for many years now. Along the way she has benefited from the wisdom, friendship and professionalism of many collaborators: Davide Bianchini, 
Olivier Blondeau-Fournier, Pasquale Calabrese, Luca Capizzi, John L. Cardy, Cecilia De Fazio, Benjamin Doyon, D\'avid X. Horv\'ath, Mat\'e Lencs\'es, Emanuele Levi,
Michele Mazzoni, Stefano Negro, Francesco Ravanini, Fabio Sailis, Istv\'an M. Sz\'ecs\'enyi and Jacopo Viti. She is particularly grateful to Pasquale Calabrese, Luca Capizzi, and  D\'avid X. Horv\'ath, with whom she took her first steps into the world of symmetry resolved entanglement measures. 

\bibliography{Ref}

\providecommand{\href}[2]{#2}\begingroup\raggedright\begin{thebibliography}{100}

\bibitem{CallanW94}
C.~J. Callan and F.~Wilczek, {\it {On geometric entropy}},  {\em Phys. Lett.} {\bf B333} (1994) 55--61 [\href{http://arXiv.org/abs/hep-th/9401072}{{\tt hep-th/9401072}}].

\bibitem{HolzheyLW94}
C.~Holzhey, F.~Larsen and F.~Wilczek, {\it {Geometric and renormalized entropy in conformal field theory}},  {\em Nucl. Phys. B} {\bf 424} (1994) 443--467 [\href{http://arXiv.org/abs/hep-th/9403108}{{\tt hep-th/9403108}}].

\bibitem{Calabrese:2004eu}
P.~Calabrese and J.~L. Cardy, {\it {Entanglement entropy and quantum field theory}},  {\em J. Stat. Mech.} {\bf 0406} (2004) P06002 [\href{http://arXiv.org/abs/hep-th/0405152}{{\tt hep-th/0405152}}].

\bibitem{latorre1}
G.~Vidal, J.~I. Latorre, E.~Rico and A.~Kitaev, {\it {Entanglement in Quantum Critical Phenomena}},  {\em Phys. Rev. Lett.} {\bf 90} (2003) 227902 [\href{http://arXiv.org/abs/quant-ph/0304098}{{\tt quant-ph/0304098}}].

\bibitem{Latorre2}
J.~I. Latorre, E.~Rico and G.~Vidal, {\it {Ground state entanglement in quantum spin chains}},  {\em Quant. Inf. Comput.} {\bf 4} (2004), no.~1 48--92 [\href{http://arXiv.org/abs/quant-ph/0304098}{{\tt quant-ph/0304098}}].

\bibitem{Jin}
B.-Q. Jin and V.~Korepin, {\it {Quantum spin chain, Toeplitz determinants and Fisher-Hartwig conjecture}},  {\em J. Stat. Phys.} {\bf 116} (2004) 79--95 [\href{http://arXiv.org/abs/quant-ph/0304108}{{\tt quant-ph/0304108}}].

\bibitem{latorre3}
J.~I. Latorre, C.~A. Lutken, E.~Rico and G.~Vidal, {\it {Fine grained entanglement loss along renormalization group flows}},  {\em Phys. Rev. A} {\bf 71} (2005) 034301 [\href{http://arXiv.org/abs/quant-ph/0404120}{{\tt quant-ph/0404120}}].

\bibitem{bennett}
C.~H. Bennett, H.~J. Bernstein, S.~Popescu and B.~Schumacher, {\it {Concentrating partial entanglement by local operations}},  {\em Phys. Rev. A} {\bf 53} (1996) 2046--2052 [\href{http://arXiv.org/abs/quant-ph/9511030}{{\tt quant-ph/9511030}}].

\bibitem{Plenio}
M.~B. Plenio and S.~Virmani, {\it {An introduction to entanglement measures}},  {\em Quant. Inf. Comput.} {\bf 7} (2007), no.~1-2 001--051 [\href{http://arXiv.org/abs/quant-ph/0504163}{{\tt quant-ph/0504163}}].

\bibitem{destillable}
E.~M. Rains, {\it {A Rigorous treatment of distillable entanglement}},  {\em Phys. Rev. A} {\bf 60} (1999) 173 [\href{http://arXiv.org/abs/quant-ph/9809078}{{\tt quant-ph/9809078}}].

\bibitem{GS}
M.~Goldstein and E.~Sela, {\it {Symmetry-Resolved Entanglement in Many-Body Systems}},  {\em Phys. Rev. Lett.} {\bf 120} (2018) 200602 [\href{http://arXiv.org/abs/1711.09418}{{\tt 1711.09418}}].

\bibitem{german3}
J.~C. Xavier, F.~C. Alcaraz and G.~Sierra, {\it Equipartition of the entanglement entropy},  {\em Phys. Rev. B} {\bf 98} (2018), no.~4 [\href{http://arXiv.org/abs/1804.06357}{{\tt 1804.06357}}].

\bibitem{entropy}
J.~L. Cardy, O.~A. Castro-Alvaredo and B.~Doyon, {\it {Form factors of branch-point twist fields in quantum integrable models and entanglement entropy}},  {\em J. Stat. Phys.} {\bf 130} (2008) 129--168 [\href{http://arXiv.org/abs/0706.3384}{{\tt 0706.3384}}].

\bibitem{benext}
B.~Doyon, {\it {Bi-partite entanglement entropy in massive two-dimensional quantum field theory}},  {\em Phys. Rev. Lett.} {\bf 102} (2009) 031602 [\href{http://arXiv.org/abs/0803.1999}{{\tt 0803.1999}}].

\bibitem{Islam}
R.~Islam, R.~Ma, P.~M. Preiss, M.~Eric~Tai, A.~Lukin, M.~Rispoli and M.~Greiner, {\it Measuring entanglement entropy in a quantum many-body system},  {\em Nature} {\bf 528} (2015), no.~7580 77--83 [\href{http://arXiv.org/abs/1509.01160}{{\tt 1509.01160}}].

\bibitem{expSRE1}
A.~Neven {\em et.~al.}, {\it {Symmetry-resolved entanglement detection using partial transpose moments}},  {\em npj Quantum Inf.} {\bf 7} (2021) 152 [\href{http://arXiv.org/abs/2103.07443}{{\tt 2103.07443}}].

\bibitem{expSRE2}
A.~Rath, V.~Vitale, S.~Murciano, M.~Votto, J.~Dubail, R.~Kueng, C.~Branciard, P.~Calabrese and B.~Vermersch, {\it {Entanglement Barrier and its Symmetry Resolution: Theory and Experimental Observation}},  {\em PRX Quantum} {\bf 4} (2023), no.~1 010318 [\href{http://arXiv.org/abs/2209.04393}{{\tt 2209.04393}}].

\bibitem{expSRE3}
V.~Vitale, A.~Elben, R.~Kueng, A.~Neven, J.~Carrasco, B.~Kraus, P.~Zoller, P.~Calabrese, B.~Vermersch and M.~Dalmonte, {\it {Symmetry-resolved dynamical purification in synthetic quantum matter}},  {\em SciPost Phys.} {\bf 12} (2022) 106 [\href{http://arXiv.org/abs/2101.07814}{{\tt 2101.07814}}].

\bibitem{expSRE4}
D.~Azses, R.~Haenel, Y.~Naveh, R.~Raussendorf, E.~Sela and E.~G. Dalla~Torre, {\it {Identification of Symmetry-Protected Topological States on Noisy Quantum Computers}},  {\em Phys. Rev. Lett.} {\bf 125} (2020), no.~12 120502 [\href{http://arXiv.org/abs/2002.04620}{{\tt 2002.04620}}].

\bibitem{Lukin_2019}
A.~Lukin, M.~Rispoli, R.~Schittko, M.~E. Tai, A.~M. Kaufman, S.~Choi, V.~Khemani, J.~Léonard and M.~Greiner, {\it Probing entanglement in a many-body–localized system},  {\em Science} {\bf 364} (2019), no.~6437 256–260 [\href{http://arXiv.org/abs/1805.09819}{{\tt 1805.09819}}].

\bibitem{Wiseman}
H.~M. Wiseman and J.~A. Vaccaro, {\it {Entanglement of Indistinguishable Particles Shared between Two Parties}},  {\em Phys. Rev. Lett.} {\bf 91} (2003) 097902 [\href{http://arXiv.org/abs/quant-ph/0210002}{{\tt quant-ph/0210002}}].

\bibitem{FC}
M.~Fagotti and P.~Calabrese, {\it {Evolution of entanglement entropy following a quantum quench: Analytic results for the $XY$ chain in a transverse magnetic field}},  {\em Phys. Rev. A} {\bf 78} (2008) 010306 [\href{http://arXiv.org/abs/0804.3559}{{\tt 0804.3559}}].

\bibitem{Calabrese:2005in}
P.~Calabrese and J.~L. Cardy, {\it Evolution of entanglement entropy in one-dimensional systems},  {\em J. Stat. Mech.} {\bf 0504} (2005) P010 [\href{http://arXiv.org/abs/cond-mat/0503393}{{\tt cond-mat/0503393}}].

\bibitem{quench}
P.~Calabrese and J.~L. Cardy, {\it {Time-dependence of correlation functions following a quantum quench}},  {\em Phys. Rev. Lett.} {\bf 96} (2006) 136801 [\href{http://arXiv.org/abs/cond-mat/0601225}{{\tt cond-mat/0601225}}].

\bibitem{quench2}
P.~Calabrese and J.~Cardy, {\it {Quantum Quenches in Extended Systems}},  {\em J. Stat. Mech.} {\bf 0706} (2007) P06008 [\href{http://arXiv.org/abs/0704.1880}{{\tt 0704.1880}}].

\bibitem{quench3}
P.~Calabrese and J.~Cardy, {\it {Entanglement and correlation functions following a local quench: a conformal field theory approach}},  {\em J. Stat. Mech.} {\bf 0710} (2007), no.~10 P10004 [\href{http://arXiv.org/abs/0708.3750}{{\tt 0708.3750}}].

\bibitem{quenchesCC}
P.~Calabrese and J.~Cardy, {\it {Quantum quenches in 1+1 dimensional conformal field theories}},  {\em J. Stat. Mech.} {\bf 1606} (2016), no.~6 064003 [\href{http://arXiv.org/abs/1603.02889}{{\tt 1603.02889}}].

\bibitem{AC}
V.~Alba and P.~Calabrese, {\it Entanglement and thermodynamics after a quantum quench in integrable systems},  {\em PNAS} {\bf 114} (2017), no.~30 7947--7951 [\href{http://arXiv.org/abs/https://www.pnas.org/content/114/30/7947.full.pdf}{{\tt https://www.pnas.org/content/114/30/7947.full.pdf}}].

\bibitem{AC2}
V.~Alba and P.~Calabrese, {\it {Entanglement dynamics after quantum quenches in generic integrable systems}},  {\em SciPost Phys.} {\bf 4} (2018), no.~3 017 [\href{http://arXiv.org/abs/1712.07529}{{\tt 1712.07529}}].

\bibitem{BaBe}
O.~Babelon and D.~Bernard, {\it {From form-factors to correlation functions: The Ising model}},  {\em Phys. Lett. B} {\bf 288} (1992) 113--120 [\href{http://arXiv.org/abs/hep-th/9206003}{{\tt hep-th/9206003}}].

\bibitem{BeLe}
D.~Bernard and A.~LeClair, {\it {Differential equations for Sine-Gordon correlation functions at the free fermion point}},  {\em Nucl. Phys. B} {\bf 426} (1994) 534--558 [\href{http://arXiv.org/abs/hep-th/9402144}{{\tt hep-th/9402144}}].

\bibitem{BeLe0}
D.~Bernard and A.~Leclair, {\it {Errata for: Differential equations for Sine-Gordon correlation functions at the free fermion point}},  {\em Nucl. Phys. B} {\bf 498} (1997) 619--622 [\href{http://arXiv.org/abs/hep-th/9703055}{{\tt hep-th/9703055}}].

\bibitem{Yurov}
V.~P. Yurov and A.~B. Zamolodchikov, {\it {Correlation functions of integrable 2-D models of relativistic field theory. Ising model}},  {\em Int. J. Mod. Phys. A} {\bf 6} (1991) 3419--3440.

\bibitem{Kniz}
V.~G. Knizhnik, {\it {Analytic Fields on Riemann Surfaces. 2}},  {\em Commun. Math. Phys.} {\bf 112} (1987) 567--590.

\bibitem{orbifold}
L.~Dixon, D.~Friedan, E.~Martinec and S.~Shenker, {\it The conformal field theory of orbifolds},  {\em Nuclear Physics B} {\bf 282} (1987) 13--73.

\bibitem{CDL}
O.~A. Castro-Alvaredo, B.~Doyon and E.~Levi, {\it {Arguments towards a c-theorem from branch-point twist fields}},  {\em J. Phys. A} {\bf 44} (2011) 492003 [\href{http://arXiv.org/abs/1107.4280}{{\tt 1107.4280}}].

\bibitem{Levi}
E.~Levi, {\it {Composite branch-point twist fields in the Ising model and their expectation values}},  {\em J. Phys. A} {\bf 45} (2012) 275401 [\href{http://arXiv.org/abs/1204.1192}{{\tt 1204.1192}}].

\bibitem{BCDLR}
D.~Bianchini, O.~A. Castro-Alvaredo, B.~Doyon, E.~Levi and F.~Ravanini, {\it {Entanglement Entropy of Non Unitary Conformal Field Theory}},  {\em J. Phys. A} {\bf 48} (2015), no.~4 04FT01 [\href{http://arXiv.org/abs/1405.2804}{{\tt 1405.2804}}].

\bibitem{bcd15}
D.~Bianchini, O.~A. Castro-Alvaredo and B.~Doyon, {\it {Entanglement Entropy of Non-Unitary Integrable Quantum Field Theory}},  {\em Nucl. Phys. B} {\bf 896} (2015) 835--880 [\href{http://arXiv.org/abs/1502.03275}{{\tt 1502.03275}}].

\bibitem{Zamolodchikov:1977yy}
A.~B. Zamolodchikov, {\it {Exact $S$-matrix of quantum solitons of the sine-Gordon model}},  {\em JETP Lett.} {\bf 25} (1977) 468.

\bibitem{ourPartIII}
L.~Capizzi, M.~Mazzoni and O.~A. Castro-Alvaredo, {\it {Symmetry resolved entanglement of excited states in quantum field theory. Part III. Bosonic and fermionic negativity}},  {\em JHEP} {\bf 06} (2023) 074 [\href{http://arXiv.org/abs/2302.02666}{{\tt 2302.02666}}].

\bibitem{BCD}
D.~Bianchini, O.~Castro-Alvaredo and B.~Doyon, {\it {Entanglement Entropy of Non-Unitary Integrable Quantum Field Theory}},  {\em Nucl. Phys.} {\bf B896} (2015) 835--880 [\href{http://arXiv.org/abs/1502.03275}{{\tt 1502.03275}}].

\bibitem{solitonbook}
R.~Rajaraman, {\em {Solitons and Instantons: An Introduction to Solitons and Instantons in Quantum Field Theory}}.
\newblock North-Holland Publishing Company, Amsterdam, The Netherlands, 1982.

\bibitem{Laflorencie_2014}
N.~Laflorencie and S.~Rachel, {\it {Spin-resolved entanglement spectroscopy of critical spin chains and Luttinger liquids}},  {\em J. Stat. Mech.} {\bf 2014} (2014), no.~11 P11013 [\href{http://arXiv.org/abs/1407.3779}{{\tt 1407.3779}}].

\bibitem{Bonsignori_2019}
R.~Bonsignori, P.~Ruggiero and P.~Calabrese, {\it {Symmetry resolved entanglement in free fermionic systems}},  {\em J. Phys. A} {\bf 52} (2019), no.~47 475302 [\href{http://arXiv.org/abs/1907.02084}{{\tt 1907.02084}}].

\bibitem{Capizzi_2020}
L.~Capizzi, P.~Ruggiero and P.~Calabrese, {\it {Symmetry resolved entanglement entropy of excited states in a CFT}},  {\em J. Stat. Mech.} {\bf 2020} (2020), no.~7 073101 [\href{http://arXiv.org/abs/2003.04670}{{\tt 2003.04670}}].

\bibitem{Bonsignori:2020laa}
R.~Bonsignori and P.~Calabrese, {\it {Boundary effects on symmetry resolved entanglement}},  {\em J. Phys. A} {\bf 54} (2021), no.~1 015005 [\href{http://arXiv.org/abs/2009.08508}{{\tt 2009.08508}}].

\bibitem{Estienne:2020txv}
B.~Estienne, Y.~Ikhlef and A.~Morin-Duchesne, {\it {Finite-size corrections in critical symmetry-resolved entanglement}},  {\em SciPost Phys.} {\bf 10} (2021), no.~3 054 [\href{http://arXiv.org/abs/2010.10515}{{\tt 2010.10515}}].

\bibitem{di2023boundary}
G.~Di~Giulio, R.~Meyer, C.~Northe, H.~Scheppach and S.~Zhao, {\it On the boundary conformal field theory approach to symmetry-resolved entanglement},  {\em SciPost Physics Core} {\bf 6} (2023), no.~3 049.

\bibitem{Calabrese:2021wvi}
P.~Calabrese, J.~Dubail and S.~Murciano, {\it {Symmetry-resolved entanglement entropy in Wess-Zumino-Witten models}},  {\em JHEP} {\bf 10} (2021) 067 [\href{http://arXiv.org/abs/2106.15946}{{\tt 2106.15946}}].

\bibitem{Sierra_2024}
P.~Saura-Bastida, A.~Das, G.~Sierra and J.~Molina-Vilaplana, {\it {Categorical-Symmetry Resolved Entanglement in CFT}},  {\em arXiv} (2024) [\href{http://arXiv.org/abs/2402.06322}{{\tt 2402.06322}}].

\bibitem{Cornfeld:2018wbg}
E.~Cornfeld, M.~Goldstein and E.~Sela, {\it {Imbalance entanglement: Symmetry decomposition of negativity}},  {\em Phys. Rev. A} {\bf 98} (2018), no.~3 032302 [\href{http://arXiv.org/abs/1804.00632}{{\tt 1804.00632}}].

\bibitem{Murciano:2021djk}
S.~Murciano, R.~Bonsignori and P.~Calabrese, {\it {Symmetry decomposition of negativity of massless free fermions}},  {\em SciPost Phys.} {\bf 10} (2021), no.~5 111 [\href{http://arXiv.org/abs/2102.10054}{{\tt 2102.10054}}].

\bibitem{Chen:2021pls}
H.-H. Chen, {\it {Symmetry decomposition of relative entropies in conformal field theory}},  {\em JHEP} {\bf 07} (2021) 084 [\href{http://arXiv.org/abs/2104.03102}{{\tt 2104.03102}}].

\bibitem{Capizzi:2021zga}
L.~Capizzi and P.~Calabrese, {\it {Symmetry resolved relative entropies and distances in conformal field theory}},  {\em JHEP} {\bf 10} (2021) 195 [\href{http://arXiv.org/abs/2105.08596}{{\tt 2105.08596}}].

\bibitem{Parez:2022sgc}
G.~Parez, {\it {Symmetry-resolved R\'enyi fidelities and quantum phase transitions}},  {\em Phys. Rev. B} {\bf 106} (2022), no.~23 235101 [\href{http://arXiv.org/abs/2208.09457}{{\tt 2208.09457}}].

\bibitem{Murciano:2022lsw}
S.~Murciano, P.~Calabrese and L.~Piroli, {\it {Symmetry-resolved Page curves}},  {\em Phys. Rev. D} {\bf 106} (2022), no.~4 046015 [\href{http://arXiv.org/abs/2206.05083}{{\tt 2206.05083}}].

\bibitem{Yin:2022toc}
C.~Yin and Z.~Liu, {\it {Universal Entanglement and Correlation Measure in Two-Dimensional Conformal Field Theories}},  {\em Phys. Rev. Lett.} {\bf 130} (2023), no.~13 131601 [\href{http://arXiv.org/abs/2211.11952}{{\tt 2211.11952}}].

\bibitem{Berthiere:2023gkx}
C.~Berthiere and G.~Parez, {\it {Reflected entropy and computable cross-norm negativity: Free theories and symmetry resolution}},  {\em Phys. Rev. D} {\bf 108} (2023), no.~5 054508 [\href{http://arXiv.org/abs/2307.11009}{{\tt 2307.11009}}].

\bibitem{Bruno:2023tez}
A.~Bruno, F.~Ares, S.~Murciano and P.~Calabrese, {\it {Symmetry resolution of the computable cross-norm negativity of two disjoint intervals in the massless Dirac field theory}},  {\em JHEP} {\bf 02} (2024) 009 [\href{http://arXiv.org/abs/2312.02926}{{\tt 2312.02926}}].

\bibitem{Parez:2021pgq}
G.~Parez, R.~Bonsignori and P.~Calabrese, {\it {Exact quench dynamics of symmetry resolved entanglement in a free fermion chain}},  {\em J. Stat. Mech.} {\bf 2109} (2021) 093102 [\href{http://arXiv.org/abs/2106.13115}{{\tt 2106.13115}}]. [Erratum: J.Stat.Mech. 2212, 129901 (2022)].

\bibitem{Ares:2022gjb}
F.~Ares, P.~Calabrese, G.~Di~Giulio and S.~Murciano, {\it {Multi-charged moments of two intervals in conformal field theory}},  {\em JHEP} {\bf 09} (2022) 051 [\href{http://arXiv.org/abs/2206.01534}{{\tt 2206.01534}}].

\bibitem{Gaur:2023yru}
H.~Gaur and U.~A. Yajnik, {\it {Multi-charged moments and symmetry-resolved R\'enyi entropy of free compact boson for multiple disjoint intervals}},  {\em JHEP} {\bf 01} (2024) 042 [\href{http://arXiv.org/abs/2310.14186}{{\tt 2310.14186}}].

\bibitem{Ares:2022hdh}
F.~Ares, S.~Murciano and P.~Calabrese, {\it {Symmetry-resolved entanglement in a long-range free-fermion chain}},  {\em J. Stat. Mech.} {\bf 2206} (2022), no.~6 063104 [\href{http://arXiv.org/abs/2202.05874}{{\tt 2202.05874}}].

\bibitem{Horvath:2020vzs}
D.~X. Horv\'ath and P.~Calabrese, {\it {Symmetry resolved entanglement in integrable field theories via form factor bootstrap}},  {\em JHEP} {\bf 11} (2020) 131 [\href{http://arXiv.org/abs/2008.08553}{{\tt 2008.08553}}].

\bibitem{Horvath:2021fks}
D.~X. Horvath, L.~Capizzi and P.~Calabrese, {\it {U(1) symmetry resolved entanglement in free 1+1 dimensional field theories via form factor bootstrap}},  {\em JHEP} {\bf 05} (2021) 197 [\href{http://arXiv.org/abs/2103.03197}{{\tt 2103.03197}}].

\bibitem{CTFMichele}
O.~A. Castro-Alvaredo and M.~Mazzoni, {\it {Two-point functions of composite twist fields in the Ising field theory}},  {\em J. Phys. A} {\bf 56} (2023), no.~12 124001 [\href{http://arXiv.org/abs/2301.01745}{{\tt 2301.01745}}].

\bibitem{Horvath:2021rjd}
D.~X. Horvath, P.~Calabrese and O.~A. Castro-Alvaredo, {\it {Branch Point Twist Field Form Factors in the sine-Gordon Model II: Composite Twist Fields and Symmetry Resolved Entanglement}},  {\em SciPost Phys.} {\bf 12} (2022), no.~3 088 [\href{http://arXiv.org/abs/2105.13982}{{\tt 2105.13982}}].

\bibitem{Potts}
L.~Capizzi, D.~X. Horv\'ath, P.~Calabrese and O.~A. Castro-Alvaredo, {\it {Entanglement of the 3-state Potts model via form factor bootstrap: total and symmetry resolved entropies}},  {\em JHEP} {\bf 05} (2022) 113 [\href{http://arXiv.org/abs/2108.10935}{{\tt 2108.10935}}].

\bibitem{ourPartI}
L.~Capizzi, O.~A. Castro-Alvaredo, C.~De~Fazio, M.~Mazzoni and L.~Santamar\'\i{}a-Sanz, {\it {Symmetry resolved entanglement of excited states in quantum field theory. Part I. Free theories, twist fields and qubits}},  {\em JHEP} {\bf 12} (2022) 127 [\href{http://arXiv.org/abs/2203.12556}{{\tt 2203.12556}}].

\bibitem{ourPartII}
L.~Capizzi, C.~De~Fazio, M.~Mazzoni, L.~Santamar\'\i{}a-Sanz and O.~A. Castro-Alvaredo, {\it {Symmetry resolved entanglement of excited states in quantum field theory. Part II. Numerics, interacting theories and higher dimensions}},  {\em JHEP} {\bf 12} (2022) 128 [\href{http://arXiv.org/abs/2206.12223}{{\tt 2206.12223}}].

\bibitem{Rottoli:2023jvw}
F.~Rottoli, F.~Ares, P.~Calabrese and D.~X. Horv\'ath, {\it {Entanglement entropy along a massless renormalisation flow: the tricritical to critical Ising crossover}},  {\em JHEP} {\bf 02} (2024) 053 [\href{http://arXiv.org/abs/2309.17199}{{\tt 2309.17199}}].

\bibitem{Murciano:2019wdl}
S.~Murciano, G.~Di~Giulio and P.~Calabrese, {\it {Symmetry resolved entanglement in gapped integrable systems: a corner transfer matrix approach}},  {\em SciPost Phys.} {\bf 8} (2020) 046 [\href{http://arXiv.org/abs/1911.09588}{{\tt 1911.09588}}].

\bibitem{Ryu:2006bv}
S.~Ryu and T.~Takayanagi, {\it {Holographic derivation of entanglement entropy from AdS/CFT}},  {\em Phys. Rev. Lett.} {\bf 96} (2006) 181602 [\href{http://arXiv.org/abs/hep-th/0603001}{{\tt hep-th/0603001}}].

\bibitem{Ryu:2006ef}
S.~Ryu and T.~Takayanagi, {\it {Aspects of Holographic Entanglement Entropy}},  {\em JHEP} {\bf 08} (2006) 045 [\href{http://arXiv.org/abs/hep-th/0605073}{{\tt hep-th/0605073}}].

\bibitem{Belin:2013uta}
A.~Belin, L.-Y. Hung, A.~Maloney, S.~Matsuura, R.~C. Myers and T.~Sierens, {\it {Holographic Charged Renyi Entropies}},  {\em JHEP} {\bf 12} (2013) 059 [\href{http://arXiv.org/abs/1310.4180}{{\tt 1310.4180}}].

\bibitem{Caputa:2015qbk}
P.~Caputa, M.~Nozaki and T.~Numasawa, {\it {Charged Entanglement Entropy of Local Operators}},  {\em Phys. Rev. D} {\bf 93} (2016), no.~10 105032 [\href{http://arXiv.org/abs/1512.08132}{{\tt 1512.08132}}].

\bibitem{Zhao:2020qmn}
S.~Zhao, C.~Northe and R.~Meyer, {\it {Symmetry-resolved entanglement in AdS$_{3}$/CFT$_{2}$ coupled to U(1) Chern-Simons theory}},  {\em JHEP} {\bf 07} (2021) 030 [\href{http://arXiv.org/abs/2012.11274}{{\tt 2012.11274}}].

\bibitem{Weisenberger:2021eby}
K.~Weisenberger, S.~Zhao, C.~Northe and R.~Meyer, {\it {Symmetry-resolved entanglement for excited states and two entangling intervals in AdS$_{3}$/CFT$_{2}$}},  {\em JHEP} {\bf 12} (2021) 104 [\href{http://arXiv.org/abs/2108.09210}{{\tt 2108.09210}}].

\bibitem{Baiguera:2022sao}
S.~Baiguera, L.~Bianchi, S.~Chapman and D.~A. Galante, {\it {Shape deformations of charged R\'enyi entropies from holography}},  {\em JHEP} {\bf 06} (2022) 068 [\href{http://arXiv.org/abs/2203.15028}{{\tt 2203.15028}}].

\bibitem{Zhao:2022wnp}
S.~Zhao, C.~Northe, K.~Weisenberger and R.~Meyer, {\it {Charged moments in W$_{3}$ higher spin holography}},  {\em JHEP} {\bf 05} (2022) 166 [\href{http://arXiv.org/abs/2202.11111}{{\tt 2202.11111}}].

\bibitem{FG}
N.~Feldman and M.~Goldstein, {\it {Dynamics of Charge-Resolved Entanglement after a Local Quench}},  {\em Phys. Rev. B} {\bf 100} (2019), no.~23 235146 [\href{http://arXiv.org/abs/1905.10749}{{\tt 1905.10749}}].

\bibitem{Fraenkel:2019ykl}
S.~Fraenkel and M.~Goldstein, {\it {Symmetry resolved entanglement: Exact results in 1D and beyond}},  {\em J. Stat. Mech.} {\bf 2003} (2020), no.~3 033106 [\href{http://arXiv.org/abs/1910.08459}{{\tt 1910.08459}}].

\bibitem{Murciano:2020lqq}
S.~Murciano, P.~Ruggiero and P.~Calabrese, {\it {Symmetry resolved entanglement in two-dimensional systems via dimensional reduction}},  {\em J. Stat. Mech.} {\bf 2008} (2020) 083102 [\href{http://arXiv.org/abs/2003.11453}{{\tt 2003.11453}}].

\bibitem{barghathi2018renyi}
H.~Barghathi, C.~Herdman and A.~Del~Maestro, {\it R{\'e}nyi generalization of the accessible entanglement entropy},  {\em Phys. Rev. Lett.} {\bf 121} (2018), no.~15 150501 [\href{http://arXiv.org/abs/1804.01114}{{\tt 1804.01114}}].

\bibitem{Barghathi:2019oxr}
H.~Barghathi, E.~Casiano-Diaz and A.~Del~Maestro, {\it {Operationally accessible entanglement of one-dimensional spinless fermions}},  {\em Phys. Rev. A} {\bf 100} (2019), no.~2 022324 [\href{http://arXiv.org/abs/1905.03312}{{\tt 1905.03312}}].

\bibitem{Tan_2020}
M.~T. Tan and S.~Ryu, {\it {Particle number fluctuations, Rényi entropy, and symmetry-resolved entanglement entropy in a two-dimensional Fermi gas from multidimensional bosonization}},  {\em Phys. Rev. B} {\bf 101} (2020), no.~23 [\href{http://arXiv.org/abs/1911.0145}{{\tt 1911.0145}}].

\bibitem{ESSLER_ghd_review}
F.~H. Essler, {\it {A short introduction to Generalized Hydrodynamics}},  {\em Physica A} (2022) 127572 [\href{http://arXiv.org/abs/2306.17072}{{\tt 2306.17072}}].

\bibitem{doyon_GHD_review}
B.~Doyon, {\it {Lecture notes on Generalised Hydrodynamics}},  {\em SciPost Phys. Lect. Notes} (2020) 18 [\href{http://arXiv.org/abs/1912.08496}{{\tt 1912.08496}}].

\bibitem{Bernard:2016nci}
D.~Bernard and B.~Doyon, {\it {Conformal field theory out of equilibrium: a review}},  {\em J. Stat. Mech.} {\bf 1606} (2016), no.~6 064005 [\href{http://arXiv.org/abs/1603.07765}{{\tt 1603.07765}}].

\bibitem{zwanzig2001nonequilibrium}
R.~Zwanzig, {\em {Nonequilibrium Statistical Mechanics}}.
\newblock Oxford University Press, Oxford, UK, 2001.

\bibitem{Caputa:2013eka}
P.~Caputa, G.~Mandal and R.~Sinha, {\it {Dynamical entanglement entropy with angular momentum and U(1) charge}},  {\em JHEP} {\bf 11} (2013) 052 [\href{http://arXiv.org/abs/1306.4974}{{\tt 1306.4974}}].

\bibitem{Parez:2020vsp}
G.~Parez, R.~Bonsignori and P.~Calabrese, {\it {Quasiparticle dynamics of symmetry-resolved entanglement after a quench: Examples of conformal field theories and free fermions}},  {\em Phys. Rev. B} {\bf 103} (2021), no.~4 L041104 [\href{http://arXiv.org/abs/2010.09794}{{\tt 2010.09794}}].

\bibitem{Fraenkel:2021ijv}
S.~Fraenkel and M.~Goldstein, {\it {Entanglement measures in a nonequilibrium steady state: Exact results in one dimension}},  {\em SciPost Phys.} {\bf 11} (2021), no.~4 085 [\href{http://arXiv.org/abs/2105.00740}{{\tt 2105.00740}}].

\bibitem{Bonsignori:2019naz}
R.~Bonsignori, P.~Ruggiero and P.~Calabrese, {\it {Symmetry resolved entanglement in free fermionic systems}},  {\em J. Phys. A} {\bf 52} (2019), no.~47 475302 [\href{http://arXiv.org/abs/1907.02084}{{\tt 1907.02084}}].

\bibitem{chen2023energy}
J.~Chen, C.~H. Chen and X.~Wang, {\it {Energy- and Symmetry-Resolved Entanglement Dynamics in Disordered Bose-Hubbard Chain}},  2023.
\newblock \href{http://arXiv.org/abs/2303.14825}{{\tt 2303.14825}}.

\bibitem{Parez:2022xur}
G.~Parez, R.~Bonsignori and P.~Calabrese, {\it {Dynamics of charge-imbalance-resolved entanglement negativity after a quench in a free-fermion model}},  {\em J. Stat. Mech.} {\bf 2205} (2022), no.~5 053103 [\href{http://arXiv.org/abs/2202.05309}{{\tt 2202.05309}}]. [Erratum: J.Stat.Mech. 2308, 089902 (2023)].

\bibitem{Scopa_2022}
S.~Scopa and D.~X. Horváth, {\it Exact hydrodynamic description of symmetry-resolved rényi entropies after a quantum quench},  {\em J. Stat. Mech.} {\bf 2022} (Aug., 2022) 083104.

\bibitem{Turkeshi:2020yxd}
X.~Turkeshi, P.~Ruggiero, V.~Alba and P.~Calabrese, {\it {Entanglement equipartition in critical random spin chains}},  {\em Phys. Rev. B} {\bf 102} (2020), no.~1 014455 [\href{http://arXiv.org/abs/2005.03331}{{\tt 2005.03331}}].

\bibitem{Kiefer_Emmanouilidis_2020}
M.~Kiefer-Emmanouilidis, R.~Unanyan, M.~Fleischhauer and J.~Sirker, {\it {Evidence for Unbounded Growth of the Number Entropy in Many-Body Localized Phases}},  {\em Phys. Rev. Lett.} {\bf 124} (2020), no.~24 [\href{http://arXiv.org/abs/2003.04849}{{\tt 2003.04849}}].

\bibitem{Kiefer_Emmanouilidis_2021}
M.~Kiefer-Emmanouilidis, R.~Unanyan, M.~Fleischhauer and J.~Sirker, {\it Unlimited growth of particle fluctuations in many-body localized phases},  {\em Annals of Physics} {\bf 435} (2021) 168481 [\href{http://arXiv.org/abs/2012.12436}{{\tt 2012.12436}}].

\bibitem{Monkman:2020ycn}
K.~Monkman and J.~Sirker, {\it {Operational entanglement of symmetry-protected topological edge states}},  {\em Phys. Rev. Res.} {\bf 2} (2020), no.~4 043191 [\href{http://arXiv.org/abs/2005.13026}{{\tt 2005.13026}}].

\bibitem{Azses:2020wfx}
D.~Azses and E.~Sela, {\it {Symmetry-resolved entanglement in symmetry-protected topological phases}},  {\em Phys. Rev. B} {\bf 102} (2020), no.~23 235157 [\href{http://arXiv.org/abs/2008.09332}{{\tt 2008.09332}}].

\bibitem{Monkman:2023hup}
K.~Monkman and J.~Sirker, {\it {Symmetry-resolved entanglement: general considerations, calculation from correlation functions, and bounds for symmetry-protected topological phases}},  {\em J. Phys. A} {\bf 56} (2023), no.~49 495001 [\href{http://arXiv.org/abs/2307.05820}{{\tt 2307.05820}}].

\bibitem{Monk}
K.~Monkman and J.~Sirker, {\it Symmetry-resolved entanglement of ${C}_{2}$-symmetric topological insulators},  {\em Phys. Rev. B} {\bf 107} (2023) 125108.

\bibitem{Horvath:2023xoh}
D.~X. Horvath, S.~Fraenkel, S.~Scopa and C.~Rylands, {\it {Charge-resolved entanglement in the presence of topological defects}},  {\em Phys. Rev. B} {\bf 108} (2023), no.~16 165406 [\href{http://arXiv.org/abs/2306.15532}{{\tt 2306.15532}}].

\bibitem{Cornfeld:2018wtp}
E.~Cornfeld, L.~A. Landau, K.~Shtengel and E.~Sela, {\it {Entanglement spectroscopy of non-Abelian anyons: Reading off quantum dimensions of individual anyons}},  {\em Phys. Rev. B} {\bf 99} (2019), no.~11 115429 [\href{http://arXiv.org/abs/1810.01853}{{\tt 1810.01853}}].

\bibitem{nonlocalqft}
R.~Pirmoradian and M.~R. Tanhayi, {\it {Symmetry-Resolved Entanglement Entropy for Local and Non-local QFTs}},  \href{http://arXiv.org/abs/2311.00494}{{\tt 2311.00494}}.

\bibitem{DMS}
P.~Di~Francesco, P.~Mathieu and D.~Senechal, {\em {Conformal Field Theory}}.
\newblock Springer, New York, USA, 1997.

\bibitem{Zamolodchikov:1989zs}
A.~B. Zamolodchikov, {\it {Integrable field theory from conformal field theory}},  {\em Adv. Stud. Pure Math.} {\bf 19} (1989) 641--674.

\bibitem{KW}
M.~Karowski and P.~Weisz, {\it {Exact S matrices and form-factors in (1+1)-dimensional field theoretic models with soliton behavior}},  {\em Nucl. Phys.} {\bf B139} (1978) 455--476.

\bibitem{smirnovbook}
F.~Smirnov, {\it Form factors in completely integrable models of quantum field theory},  {\em Adv. Series in Math. Phys.} {\bf 14} (1992) World Scientific, Singapore.

\bibitem{hastings}
M.~B. Hastings, {\it Entropy and entanglement in quantum ground states},  {\em Phys. Rev. B} {\bf 76} (2007) 035114 [\href{http://arXiv.org/abs/cond-mat/0701055}{{\tt cond-mat/0701055}}].

\bibitem{ZA}
A.~Zamolodchikov and A.~Zamolodchikov, {\it {Factorized S-matrices in two-dimensions as the exact solutions of certain relativistic quantum field models}},  {\em Annals Phys.} {\bf 120} (1979) 253--291.

\bibitem{FA}
L.~D. Faddeev, {\em {Quantum Completely Integrable Models in Field Theory}}, pp.~187--235.
\newblock 1995.

\bibitem{Castro-Alvaredo:2008usl}
O.~A. Castro-Alvaredo and B.~Doyon, {\it {Bi-partite entanglement entropy in integrable models with backscattering}},  {\em J. Phys. A} {\bf 41} (2008) 275203 [\href{http://arXiv.org/abs/0802.4231}{{\tt 0802.4231}}].

\bibitem{Castro-Alvaredo:2021jjl}
O.~A. Castro-Alvaredo and D.~X. Horvath, {\it {Branch point twist field form factors in the sine-Gordon model I: Breather fusion and entanglement dynamics}},  {\em SciPost Phys.} {\bf 10} (2021), no.~6 132 [\href{http://arXiv.org/abs/2103.08492}{{\tt 2103.08492}}].

\bibitem{KK}
A.~Koubek and G.~Mussardo, {\it {On the operator content of the sinh-Gordon model}},  {\em Phys. Lett.} {\bf B311} (1993) 193--201 [\href{http://arXiv.org/abs/hep-th/9306044}{{\tt hep-th/9306044}}].

\bibitem{Luky1}
S.~L. Lukyanov, {\it {Form factors of exponential fields in the sine-Gordon model}},  {\em Mod. Phys. Lett.} {\bf A12} (1997) 2543--2550 [\href{http://arXiv.org/abs/hep-th/9703190}{{\tt hep-th/9703190}}].

\bibitem{Takacs:2008zz}
G.~Takacs, {\it {Form-factors of boundary exponential operators in the sinh-Gordon model}},  {\em Nucl. Phys. B} {\bf 801} (2008), no.~3 187--206 [\href{http://arXiv.org/abs/0801.0962}{{\tt 0801.0962}}].

\bibitem{nexttonext}
O.~A. Castro-Alvaredo and B.~Doyon, {\it {Bi-partite entanglement entropy in massive QFT with a boundary: the Ising model}},  {\em J. Stat. Phys.} {\bf 134} (2009) 105--145 [\href{http://arXiv.org/abs/0810.0219}{{\tt 0810.0219}}].

\bibitem{mussardobook}
G.~Mussardo, {\em {Statistical Field Theory: An Introduction to Exactly Solved Models in Statistical Physics}}.
\newblock Oxford University Press, Oxford, UK, 2009.

\bibitem{next}
B.~Doyon, {\it {Bi-partite entanglement entropy in massive two-dimensional quantum field theory}},  {\em Phys. Rev. Lett.} {\bf 102} (2009) 031602 [\href{http://arXiv.org/abs/0803.1999}{{\tt 0803.1999}}].

\bibitem{german1}
F.~C. Alcaraz, M.~I. Berganza and G.~Sierra, {\it {Entanglement of low-energy excitations in Conformal Field Theory}},  {\em Phys. Rev. Lett.} {\bf 106} (2011) 201601 [\href{http://arXiv.org/abs/1101.2881}{{\tt 1101.2881}}].

\bibitem{german2}
M.~I. Berganza, F.~C. Alcaraz and G.~Sierra, {\it {Entanglement of excited states in critical spin chains}},  {\em J. Stat. Mech.} {\bf 1201} (2012) P01016 [\href{http://arXiv.org/abs/1109.5673}{{\tt 1109.5673}}].

\bibitem{Vincenzo2}
V.~Alba, M.~Fagotti and P.~Calabrese, {\it Entanglement entropy of excited states},  {\em J. Stat. Mech.} {\bf 2009} (2009), no.~10 P10020 [\href{http://arXiv.org/abs/0909.1999}{{\tt 0909.1999}}].

\bibitem{Vincenzo}
J.~M\"olter, T.~Barthel, U.~Schollw\"ock and V.~Alba, {\it Bound states and entanglement in the excited states of quantum spin chains},  {\em J. Stat. Mech.} {\bf 2014} (2014), no.~10 P10029 [\href{http://arXiv.org/abs/1407.0066}{{\tt 1407.0066}}].

\bibitem{excited}
O.~A. Castro-Alvaredo, C.~De~Fazio, B.~Doyon and I.~M. Sz\'ecs\'enyi, {\it {Entanglement Content of Quasiparticle Excitations}},  {\em Phys. Rev. Lett.} {\bf 121} (2018) 170602 [\href{http://arXiv.org/abs/1805.04948}{{\tt 1805.04948}}].

\bibitem{excited1}
O.~A. Castro-Alvaredo, C.~De~Fazio, B.~Doyon and I.~M. Sz\'ecs\'enyi, {\it {Entanglement content of quantum particle excitations. Part I. Free field theory}},  {\em JHEP} {\bf 10} (2018) 039 [\href{http://arXiv.org/abs/1806.03247}{{\tt 1806.03247}}].

\bibitem{excited2}
O.~A. Castro-Alvaredo, C.~De~Fazio, B.~Doyon and I.~M. Sz\'ecs\'enyi, {\it {Entanglement content of quantum particle excitations. Part II. Disconnected regions and logarithmic negativity}},  {\em JHEP} {\bf 11} (2019) 058 [\href{http://arXiv.org/abs/1904.01035}{{\tt 1904.01035}}].

\bibitem{excited3}
O.~A. Castro-Alvaredo, C.~De~Fazio, B.~Doyon and I.~M. Sz\'ecs\'enyi, {\it {Entanglement Content of Quantum Particle Excitations III. Graph Partition Functions}},  {\em J. Math. Phys.} {\bf 60} (2019), no.~8 082301 [\href{http://arXiv.org/abs/1904.02615}{{\tt 1904.02615}}].

\bibitem{Zhang1}
J.~Zhang and M.~A. Rajabpour, {\it {Excited state R\'enyi entropy and subsystem distance in two-dimensional non-compact bosonic theory. Part I. Single-particle states}},  {\em JHEP} {\bf 12} (2020) 160 [\href{http://arXiv.org/abs/2009.00719}{{\tt 2009.00719}}].

\bibitem{Zhang_2021a}
J.~Zhang and M.~A. Rajabpour, {\it {Universal Rényi entanglement entropy of quasiparticle excitations}},  {\em Europhysics Letters} {\bf 135} (2021), no.~6 60001 [\href{http://arXiv.org/abs/2010.13973}{{\tt 2010.13973}}].

\bibitem{Zhang2}
J.~Zhang and M.~A. Rajabpour, {\it {Corrections to universal R\'enyi entropy in quasiparticle excited states of quantum chains}},  {\em J. Stat. Mech.} {\bf 2109} (2021) 093101 [\href{http://arXiv.org/abs/2010.16348}{{\tt 2010.16348}}].

\bibitem{Zhang3}
J.~Zhang and M.~A. Rajabpour, {\it {Excited state R\'enyi entropy and subsystem distance in two-dimensional non-compact bosonic theory. Part II. Multi-particle states}},  {\em JHEP} {\bf 08} (2021) 106 [\href{http://arXiv.org/abs/2011.11006}{{\tt 2011.11006}}].

\bibitem{Zhang4}
J.~Zhang and M.~A. Rajabpour, {\it {Entanglement of magnon excitations in spin chains}},  {\em JHEP} {\bf 02} 072 [\href{http://arXiv.org/abs/2109.12826}{{\tt 2109.12826}}].

\bibitem{Mussardo:2021gws}
G.~Mussardo and J.~Viti, {\it {\ensuremath{\hbar}\textrightarrow{}0 limit of the entanglement entropy}},  {\em Phys. Rev. A} {\bf 105} (2022), no.~3 032404 [\href{http://arXiv.org/abs/2112.06840}{{\tt 2112.06840}}].

\bibitem{EERev2}
J.~Eisert, M.~Cramer and M.~B. Plenio, {\it {Colloquium: Area laws for the entanglement entropy}},  {\em Rev. Mod. Phys.} {\bf 82} (2010) 277--306 [\href{http://arXiv.org/abs/0808.3773}{{\tt 0808.3773}}].

\bibitem{PT1}
B.~Pozsgay and G.~Takacs, {\it {Form-factors in finite volume I: Form-factor bootstrap and truncated conformal space}},  {\em Nucl. Phys.} {\bf B788} (2008) 167--208 [\href{http://arXiv.org/abs/0706.1445}{{\tt 0706.1445}}].

\bibitem{PT2}
B.~Pozsgay and G.~Takacs, {\it {Form factors in finite volume. II. Disconnected terms and finite temperature correlators}},  {\em Nucl. Phys.} {\bf B788} (2008) 209--251 [\href{http://arXiv.org/abs/0706.3605}{{\tt 0706.3605}}].

\bibitem{doubling}
P.~Fonseca and A.~Zamolodchikov, {\it {Ward identities and integrable differential equations in the Ising field theory}},  \href{http://arXiv.org/abs/hep-th/0309228}{{\tt hep-th/0309228}}.

\bibitem{YZam}
V.~P. Yurov and A.~B. Zamolodchikov, {\it {Correlation functions of integrable 2-D models of relativistic field theory. Ising model}},  {\em Int. J. Mod. Phys.} {\bf A6} (1991) 3419--3440.

\bibitem{Ares:2022koq}
F.~Ares, S.~Murciano and P.~Calabrese, {\it {Entanglement asymmetry as a probe of symmetry breaking}},  {\em Nature Commun.} {\bf 14} (2023), no.~1 2036 [\href{http://arXiv.org/abs/2207.14693}{{\tt 2207.14693}}].

\bibitem{Capizzi:2020mio}
L.~Capizzi and M.~Mazzoni, {\it {Entanglement asymmetry in the ordered phase of many-body systems: the Ising field theory}},  {\em JHEP} {\bf 23} (2020) 144 [\href{http://arXiv.org/abs/2307.12127}{{\tt 2307.12127}}].

\bibitem{Ares:2023kcz}
F.~Ares, S.~Murciano, E.~Vernier and P.~Calabrese, {\it {Lack of symmetry restoration after a quantum quench: An entanglement asymmetry study}},  {\em SciPost Phys.} {\bf 15} (2023), no.~3 089 [\href{http://arXiv.org/abs/2302.03330}{{\tt 2302.03330}}].

\bibitem{Bertini:2023ysg}
B.~Bertini, K.~Klobas, M.~Collura, P.~Calabrese and C.~Rylands, {\it {Dynamics of charge fluctuations from asymmetric initial states}},  {\em arXiv} (2023) [\href{http://arXiv.org/abs/2306.12404}{{\tt 2306.12404}}].

\bibitem{Capizzi:2023xaf}
L.~Capizzi and V.~Vitale, {\it {A universal formula for the entanglement asymmetry of matrix product states}},  {\em arXiv} (2023) [\href{http://arXiv.org/abs/2310.01962}{{\tt 2310.01962}}].

\bibitem{Ares:2023ggj}
F.~Ares, S.~Murciano, L.~Piroli and P.~Calabrese, {\it {An entanglement asymmetry study of black hole radiation}},  {\em arXiv} (2023) [\href{http://arXiv.org/abs/2311.12683}{{\tt 2311.12683}}].

\end{thebibliography}\endgroup
\end{document}